\documentclass[trackchanges,twocolumn]{aastex701} %linenumbers,
\usepackage{graphicx}
\usepackage{txfonts}
\usepackage{xcolor}
\usepackage{hyperref}
\usepackage[normalem]{ulem}
\hypersetup{colorlinks=true,urlcolor=blue,linkcolor=blue,citecolor=blue}

\def \MSUN{\rm M_{\odot}}

\def \MFIVEC{M_{\rm 500c}}
\newcommand{\mytilde}{\raise.19ex\hbox{$\scriptstyle\sim$}}

\begin{document}

\title{Exploring the statistical properties of double radio relics in the TNG-Cluster and TNG300 simulations}

\author[orcid=0000-0002-1566-5094]{Wonki Lee}
\affiliation{Yonsei University, Department of Astronomy, Seoul, Republic of Korea}
\email[show]{wonki.lee@yonsei.ac.kr}  

\author[orcid=0000-0003-1065-9274]{Annalisa Pillepich}
\affiliation{Max-Planck-Institut f{\"u}r Astronomie, K{\"o}nigstuhl 17, D-69117 Heidelberg, Germany}
\email[]{}

\author[orcid=0000-0001-8421-5890]{Dylan Nelson}
\affiliation{Zentrum f{\"u}r Astronomie der Universit{\"a}t Heidelberg, ITA, Albert Ueberle Str. 2, D-69120 Heidelberg, Germany}
\email[]{}

\author[orcid=0000-0002-5751-3697]{Myungkook James Jee}
\affiliation{Yonsei University, Department of Astronomy, Seoul, Republic of Korea}
\affiliation{Department of Physics, University of California, Davis, One Shields Avenue, Davis, CA 95616, USA}
\email[]{}

\author[orcid=0000-0002-6766-5942]{Daisuke Nagai}
\affiliation{Department of Physics, Yale University, New Haven, CT 06520, USA}
\email[]{}

\author[orcid=0000-0002-4462-0709]{Kyle Finner}
\affiliation{IPAC, California Institute of Technology, 1200 E California Blvd., Pasadena, CA 91125, USA}
\email[]{}

\author[orcid=0000-0003-3175-2347]{John ZuHone}
\affiliation{Harvard-Smithsonian Center for Astrophysics, 60 Garden St., Cambridge, MA 02138, USA}
\email[]{}

%\altaffiliation{Las Campanas Observatory}
%\affiliation{Universidad de Chile, Department of Astronomy}
%\email{fakeemail2@google.com}

%\collaboration{all}{The Terra Mater collaboration}

\begin{abstract}

Double radio relics, pairs of diffuse radio features located on opposite sides of merging galaxy clusters, are a rare subclass of radio relics that are believed to trace merger shocks and provide valuable constraints on plasma acceleration models and merger history. With the number of known double relics growing in recent and upcoming radio surveys, statistical analyses of their properties are becoming feasible. In this study, we utilize the cosmological magnetohydrodynamics zoom-in simulations TNG-Cluster, in combination with TNG300-1, to examine the statistical properties of double radio relics. The simulated double relic pairs exhibit a wide range of luminosity ratios, broadly consistent with the observations. We find that the two relics in a given double system often differ significantly in their shock properties and magnetic field strengths. This diversity implies that the observed brightness asymmetry in the pair cannot be explained by a single factor alone, 
but instead reflects an interplay of multiple physical parameters. Nevertheless, double radio relics tend to align with the collision axis within $\mytilde30^{\circ}$ and their separation ($d_{\rm drr}$) correlates tightly with the time since collision (TSC) as ${\rm TSC~[Gyr]} = 0.52 d_{\rm drr}/R_{500\rm c} - 0.24$, allowing it to be inferred with an accuracy of $\mytilde0.2~\rm Gyr$.
With the statistical samples of simulated radio relics, we predict that low-mass clusters will constitute the dominant population of double radio relic systems detected with upcoming surveys such as SKA. These results demonstrate that double radio relics can serve as robust probes of merger dynamics and plasma acceleration, and that simulations provide critical guidance for interpreting the large samples expected from next-generation radio surveys.

% and their separation correlates with the time since collision, allowing it to be inferred with an accuracy of $\mytilde0.2~\rm Gyr$. These results suggest that the statistics of double relics can serve as useful probes of acceleration models and merger dynamics and that simulations can offer guidance for interpreting the large samples from the upcoming radio surveys.

\end{abstract}



\section{Introduction} 

Radio relics are diffuse, Mpc-scale radio features observed in the periphery of merging galaxy clusters. Their origin is widely attributed to diffusive shock acceleration (DSA), which can generate cosmic-ray electrons (CRe) along merger shocks and produce synchrotron radio emission under the intracluster magnetic fields \citep[e.g.,][]{1997MNRAS.290..577R,1998A&A...332..395E,2008A&A...486..347G,2012A&ARv..20...54F}. This interpretation is supported by the short cooling timescale of CRe of $\mytilde100\rm~Myr$ in the intracluster medium (ICM) and the uniform spectral properties observed across relics \citep[e.g.,][]{2010Sci...330..347V,2013MNRAS.435.1061P}, which together require in-situ acceleration along their Mpc-scale extent. Furthermore, the high polarization of their radio emission provides additional evidence for shock compression that orders magnetic fields along the shock front. X-ray observations \citep[e.g.,][]{2013MNRAS.433..812O,2015A&A...582A..87A,2016MNRAS.460L..84B} or the Sunyaev-Zeldovich (SZ) effect analysis \citep[e.g.,][]{2016ApJ...829L..23B} have further confirmed this scenario by identifying merger shocks coincident with radio relics.

The unique capability of radio relics to trace merger shocks establishes them as powerful probes of a cluster merger history. Unlike dark matter halos or galaxies that remain bound to the cluster potential, merger shocks are launched outward from the core of the clusters after pericenter passage, providing robust constraints on the merger geometry and the time since collision \citep[TSC hereafter;][]{2017MNRAS.470.3465B,2018ApJ...857...26H,2019MNRAS.488.5259Z}. Moreover, as relics are detected even beyond $R_{\rm 500c}$, they can offer a unique observational window into the late-phase evolution of cluster mergers, where X-ray surface brightness is too faint to trace shocks effectively \citep[e.g.,][]{2014ApJ...785....1B,2024A&A...691A..99P,2025ApJ...984...25R}. When combined with multi-wavelength observations or dedicated idealized simulations, radio relics yield crucial insights into the merger history of galaxy clusters \citep[e.g.,][]{2015MNRAS.453.1531N,2019ApJ...882...69G,2020ApJ...894...60L,2025ApJS..277...28F,2025ApJ...984...26A}.

Specifically, tight constraints on the TSC are essential for understanding cluster mergers. Thermodynamic properties, including X-ray luminosity and temperature, evolve dramatically over the Gyr-timescale merger history \citep[e.g.,][]{2001ApJ...561..621R}, while shocks and turbulence gradually develop and generate diffuse radio emissions that change with the TSC \citep[e.g.,][]{2012MNRAS.421.1868V,2013MNRAS.429.3564D,2024A&A...690A.146N}. The dark matter halo also varies with the TSC \citep[e.g.,][]{2012MNRAS.419.1338R}, potentially affecting cluster mass estimates at different merger phases \citep[e.g.,][]{2022MNRAS.509.1201C,2023ApJ...945...71L}. Radio relics thus serve as powerful probes that connect these multi-wavelength features to the merger phase. Moreover, the accurate estimate of TSC further provides a reference for using cluster mergers to test the self-interacting nature of dark matter in which spatial distribution evolves with time \citep[e.g.,][]{2017MNRAS.469.1414K,2022AAS...24013917P}.

Synchrotron radio emission from shock waves also allows us to test plasma acceleration models.
Plasma-scale simulations suggest that shock strengths in cluster environments are often too weak ($\mathcal{M}\lesssim3$) to explain the observed radio luminosity through the DSA mechanism acting on thermal plasma \citep[e.g.,][]{2020A&A...634A..64B}. 
While some relics can be explained by a wide variation in shock strength and radio emission dominated by the high Mach number tails  \citep[e.g.,][]{2015ApJ...812...49H,2021MNRAS.506..396W,2025ApJ...978..122L}, challenges remain with relics with weak shock strength from both radio and X-ray observations \citep[e.g.,][]{2024MNRAS.52710986C}, along with the $\gamma$-ray non-detection in galaxy clusters \citep[][]{2014ApJ...787...18A}.
These tensions have prompted discussions of alternative plasma acceleration models involving radio relics, including the re-acceleration of fossil plasma \citep[i.e., old suprathermal plasma seeded by past merger or AGN activity; e.g.,][]{2013MNRAS.435.1061P,2016ApJ...823...13K,2017NatAs...1E...5V,2019MNRAS.489.3905S}, upstream pre-acceleration \citep[e.g.,][]{2014ApJ...797...47G,2019ApJ...876...79K,2021ApJ...915...18H}, and multi-shock scenarios where multiple passages of weak shocks boost the efficiency \citep[e.g.,][]{2021JKAS...54..103K,2022MNRAS.509.1160I}. 

Double radio relics form a distinctive subset of radio relic systems.
These systems present a pair of radio relics, located on opposite sides of the collision center \citep[e.g.,][]{1997MNRAS.290..577R,2010Sci...330..347V,2011A&A...528A..38V,2012MNRAS.425L..36V,2014ApJ...786...49L,2014MNRAS.444.3130D,2018MNRAS.478.2218H,2021PASA...38....5D,2022ApJ...924...18L,2024MNRAS.531.3357K,2024MNRAS.52710986C}. 
Although merger shocks generally propagate in opposite directions from the collision center, only a fraction of mergers reveal both shocks as bright, extended radio relics - forming what are known as double radio relics. 
As double radio relics are generated by a pair of merger shocks launched from a single collision, the two relics provide independent constraints on a common merger, offering an optimal condition for reconstructing the merger scenario in detail \citep[e.g.,][]{2021ApJ...918...72F,2022ApJ...925...68C,2025ApJ...984...26A}. Specifically, their elongated morphology, high polarization, and location provide information about the collision axis \citep[e.g.,][]{2011A&A...533A..35V,2019ApJ...882...69G}, the time since pericenter passage \citep[e.g.,][]{2021ApJ...918...72F}, and the mass ratio of colliding subclusters \citep[e.g.,][]{2011MNRAS.418..230V}.
Their well-characterized merger history, combined with the two independent synchrotron radio features generated by the same collision, makes them powerful laboratories for studying plasma acceleration models. Despite their rarity, a few well-studied systems have served as key testbeds for shock acceleration models \citep[e.g.,][]{2012ApJ...756...97K,2016MNRAS.463.1534B,2021MNRAS.505.4762J,2025A&A...698A.271K}.

With only a small number of known double relics, earlier studies mainly focused on detailed analyses of individual systems. Statistical studies of double relics, when attempted, have largely been limited to their global properties such as total luminosity or the largest linear size (LLS) and correlations with the host mass \citep[e.g.,][]{2014MNRAS.444.3130D,2021PASA...38....5D,2025ApJ...984...24S}. Recent advances in wide field deep radio observations have significantly increased the number of detected relics \citep[e.g.,][]{2024MNRAS.531.3357K,2025MNRAS.543.1638K,2025arXiv250700133B}, expanding the sample to nearly 100 relics, including $\mytilde$30 double relic systems, some of which were previously classified as single relics \citep[e.g.,][]{2022SciA....8.7623B,2024ApJ...962...40S}. With the much larger samples expected from upcoming facilities, including the LOFAR2.0 and Square Kilometre Array \citep[SKA;][]{2020MNRAS.493.2306B}, statistical analyses of double relics will become increasingly important.

Numerical simulations play an essential role in interpreting observations of radio relics. Particle-in-cell (PIC) simulations probe plasma-scale instabilities in merger shocks and their role in producing relativistic plasma \citep[e.g.,][]{2014ApJ...797...47G,2016ApJ...823...13K}. Idealized simulations explore different combinations of merger parameters, allowing the reconstruction of merger histories that can reproduce the observed clusters and their relics \citep[e.g.,][]{2011MNRAS.418..230V,2012MNRAS.425L..76B}. Cosmological simulations, on the other hand, model the evolution of galaxy clusters within the large-scale structure, with relics forming naturally during the cluster mass assembly. These relics are typically investigated either in detail with high-resolution zoom-in simulations \citep[e.g.,][]{2021MNRAS.506..396W,2023ApJ...957L..16B} or statistically across large cosmological volumes \citep[e.g.,][]{2011ApJ...735...96S,2017MNRAS.470..240N}.
These simulations support the merger history reconstruction using radio relics with the simulated analogs in idealized \citep[e.g.,][]{2007MNRAS.380..911S,2014ApJ...787..144L,2020ApJ...894...60L,2022MNRAS.509.1201C} and cosmological simulations  \citep[e.g.,][]{2013ApJ...765...21S,2016MNRAS.459...70V,2018ApJ...857...26H,2025ApJ...978..122L}.
Each approach, however, faces limitations, either in capturing the diverse nature of radio relics with sufficient statistics or in resolving radio relics located in the cluster outskirts. Hybrid methods attempt to combine different simulation approaches to mitigate these limitations \citep[e.g.,][]{2021MNRAS.500..795D,2023ApJ...943..119H,2024arXiv241111947W}. Nevertheless, a statistical analysis of double radio relics within high-resolution cosmological simulations that incorporate realistic cluster and galaxy formation physics remains unexplored.

In \citet{2024A&A...686A..55L}, we demonstrated that the diverse morphologies of observed relics can be reproduced using TNG-Cluster \citep[][]{2024A&A...686A.157N}. The suite of cosmological magnetohydrodynamics (MHD) zoom-in simulations of massive galaxy clusters selected from a Gpc-size cosmological box enabled more direct comparison with observations through a large statistical sample of clusters \citep[e.g.,][]{2024A&A...686A..86R,2024A&A...686A.200T,2024A&A...687A.129L,2024A&A...690A..20A}. Specifically, \citet{2024A&A...686A..55L} showed that the radio relics predicted from TNG-Cluster can reproduce the diverse morphology and physical properties of observed relics, including their luminosity and size. Extending that investigation, this study focuses on the statistical properties of double radio relic pairs and their connection to the host merger histories. 
%By combining the TNG-Cluster suite with the IllustrisTNG300-1 simulation (hereafter TNG300), 
By combining the TNG-Cluster suite with the TNG300 box of the IllustrisTNG project\footnote{\url{https://www.tng-project.org/}}, 
we leverage a statistically powerful sample of simulated clusters spanning masses from $10^{14}\rm~M_{\odot}$ to $10^{15.3}\rm~M_{\odot}$. Our aim is to provide guidance for upcoming wide-area deep radio surveys, which will detect statistically significant samples of double relics, and to identify the merger and acceleration properties that can be tested with such observations.

This paper is organized as follows. In \textsection\ref{sec:method}, we describe the TNG-Cluster and TNG300 simulations and the procedure for identifying radio relics and their associated cluster mergers. The statistical properties of double radio relics are presented in \textsection\ref{sec:result}, followed by an examination of their connection with merger history in \textsection\ref{sec:RRMH}. Implications for plasma acceleration models, future prospects for upcoming surveys such as the SKA, and the role of projection effects are discussed in \textsection\ref{sec:discussion}, before we summarize our findings in \textsection\ref{sec:summary}.

The TNG-Cluster and TNG300 simulations adopt a \citet{2016A&A...594A..13P} cosmology with $h=0.6774$, $\Omega_{\rm m,0}=0.3089$, $\Omega_{\rm \Lambda,0}=0.6911$, $\Omega_{\rm b,0}=0.0486$, $\sigma_{8}=0.8159$, and $n_{\rm s}=0.9667$. 
When discussing the mass and size of clusters, $R_{\Delta\rm c}$ describes the radius where the average density becomes $\Delta$ times the critical density of the universe, and $M_{\Delta\rm c}$ is the total mass within $R_{\Delta\rm c}$.

\section{Methods}
\label{sec:method}

\subsection{TNG-Cluster and TNG300}
\label{sec:tngc}

We use TNG-Cluster and TNG300-1 (TNG300 hereafter) for our analysis.
TNG300 is a cosmological MHD simulation of $\mytilde300\rm~Mpc$ size box, representing the largest volume in the IllustrisTNG series \citep[][]{2018MNRAS.475..648P,2018MNRAS.475..676S,2018MNRAS.480.5113M,2018MNRAS.475..624N,2018MNRAS.477.1206N,2019MNRAS.490.3196P,2019ComAC...6....2N,2019MNRAS.490.3234N}. 
TNG-Clsuter, a spin-off project of IllustrisTNG, consists of cosmological MHD zoom-in simulations of 352 massive galaxy clusters sampled from a 1 Gpc parent box \citep{2024A&A...686A.157N}.
In TNG-Cluster, all halos with $M_{\rm 200c}>10^{15}\rm~\MSUN$ in a Gpc-size box at $z=0$ are re-simulated, while lower-mass halos with $M_{\rm 200c}=10^{14.3-15.0}\rm~\MSUN$ are randomly sampled, providing a nearly equal number of simulated clusters in the mass regime when combined with TNG300.

The numerical setups of the two simulations are consistent with each other, which enables coherent analysis of the large number of clusters across a wide mass range.
Both simulations solve the MHD equations using the moving-mesh code \texttt{AREPO} \citep{2010MNRAS.401..791S}, and implement the TNG galaxy formation models, which includes thermal and kinetic AGN feedback, radiative cooling, star formation, galactic winds, and metal enrichment \citep{2017MNRAS.465.3291W,2018MNRAS.473.4077P}.
These models successfully reproduce the observed properties of galaxy clusters across a range of contexts \citep[e.g.,][]{2018MNRAS.474.2073V,2018MNRAS.481.1809B,2018MNRAS.481.1950L,2021MNRAS.501.1300S,2024A&A...686A.157N,2024A&A...686A..55L,2024A&A...686A.200T,2024A&A...686A..86R,2025MNRAS.536.3200P,2025MNRAS.539.1040P}.
The evolution and amplification of magnetic fields is modeled by solving the ideal continuum MHD equations \citep{2011MNRAS.418.1392P,2013MNRAS.432..176P}. 
From an initial homogeneous seed of $10^{-14}$ comoving Gauss, magnetic fields are amplified via flux conservation in gravitational collapse, turbulence, and shear flows during structure formation and feedback process, reaching typical intracluster magnetic field strengths of $\mytilde0.1-10~\mu G$ \citep{2018MNRAS.480.5113M,2024A&A...686A.157N,2025arXiv250712517L}.

Both simulations adopt the same resolution. 
Dark matter and baryons are resolved with a mass resolution of \mbox{$m_{\rm DM}\sim6\times10^{7}~\MSUN$} and \mbox{$m_{\rm baryon}\sim10^{7}~\MSUN$}, respectively. 
The gravitational softening length of dark matter and stars is $\mytilde1.4\rm~kpc$ at $z=0$, whereas gas cells use an adaptive comoving softening length with a minimum value of $\mytilde0.4\rm~kpc$. 
This resolution enables us to accurately capture the morphology and energetics of merger shocks and their associated radio emission in the cluster outskirts, with typical cell sizes of $\mytilde10$~kpc \citep[see Appendix A of][]{2024A&A...686A..55L}.

TNG-Cluster and TNG300 include an on-the-fly shock finder algorithm \citep{2015MNRAS.446.3992S, 2016MNRAS.461.4441S}, which detects shocks by identifying converging gas flows and aligned temperature and density gradients \citep[e.g.,][]{2003ApJ...593..599R,2008ApJ...689.1063S}. Shock properties, including the Mach number ($\mathcal{M}$) and shock dissipation rate ($E_{\rm diss}$), are computed using the upstream and downstream gas properties and assigned to the local shock surface cells, defined as the gas cells exhibiting the strongest velocity convergence within each shocked region. For TNG-Cluster, we analyze gas cells located within $2R_{\rm 200c}$ of the cluster center and within the high-resolution zoom region. As the TNG-Cluster halos are uncontaminated out to radii of $3R_{\rm 200c}$ \citep{2024A&A...686A.157N}, these selection criteria allow us to combine radio relics identified from TNG-Cluster and TNG300. 

We generate and use our cluster merger catalog\footnote{\url{https://www.tng-project.org/data/cluster/}} \citep[][]{2024A&A...686A..55L}. 
The catalog identifies a system as a merging cluster when two \texttt{SUBFIND} halos \citep[hereafter subhalos;][]{2001MNRAS.328..726S} undergo their first pericenter passage. 
Hereafter, the more massive cluster is referred to as the main cluster, and the less massive cluster is referred to as the sub-cluster. 
We track the relative separation between clusters, measured at all snapshots sampled with a time step of $\mytilde100\rm~Myr$, and fit a second-order polynomial to determine the pericenter separation and the time of pericenter passage.
Then, we define the mass ratio using the maximum mass of the sub-cluster before its first pericenter passage and the mass of the main cluster at the same snapshot.
We define the total mass as $\MFIVEC$ of the main cluster at the time of the first pericenter passage, and the collision axis as the vector connecting the subhalo positions at the snapshot before and after pericenter passage.
%The total mass of the cluster merger is referred to $\MFIVEC$ of the main cluster at the time of the first pericenter passage. 
%, and the collision velocity as the maximum relative velocity during the first pericenter passage.

\subsection{Radio relic properties}
\label{sec:relics}

We follow \citet{2024A&A...686A..55L} to model the properties of double radio relics. 
We assume that cosmological shockwaves accelerate thermal plasma to the non-thermal regime by transferring a fraction of the shock-dissipated energy to CRe \citep{2007MNRAS.375...77H}. 
Radio emission is assigned to shock surface cells under the assumption that shock properties and magnetic field strengths remain approximately constant over the electron cooling timescale.
Although this implementation simplifies the downstream emission and produces relatively thin radio relics, the simulated relics successfully reproduce the diverse morphologies and overall scaling relations observed in real systems \citep{2017MNRAS.470..240N,2024A&A...686A..55L}.

We analyze eight full snapshots in the redshift range $z=[0,1]$ and compute the radio emission from clusters with $M_{\rm 500c} \geq 10^{14}\rm~M_{\odot}$ in TNG300 and TNG-Cluster. 
The radio emissivity of all shock surface cells within $2R_{\rm 200c}$ is distributed into a three-dimensional histogram with a $50\rm~kpc$ bin size, where we identify radio structures by spatially connecting the bins with radio luminosity above $10^{17}\rm~W~Hz^{-1}$. 
We use a smaller bin size of $25 \rm~kpc$, if the radio emission from the merger shock pairs are merged into a single group due to their small spatial separation.
To isolate merger shock-driven radio relics from radio features of other origins, we select the two brightest groups among the five largest-volume radio structures.
This selection scheme, which combines volume and luminosity criteria, helps distinguish merger relics from radio emission associated with AGN activity and accretion shocks, respectively \citep[e.g.,][]{2025arXiv250925314P}.
Hereafter, the brightest and second-brightest radio structures are referred to as the \textit{primary} and \textit{secondary} radio relics, respectively.

Merger shocks often exhibit multiple substructures, which can appear as distinct radio relics in the simulations. However, we do not merge these into a single structure, as observations frequently report separate relics on the same side of the cluster \citep[e.g.,][]{2017ApJ...845...81P,2018MNRAS.477..957D,2021A&A...656A.154H}.
Still, we have reclassified these double relic pairs on the same side when assessing their role in merger history reconstruction in \textsection \ref{sec:RRMH}. 
Moreover, although our volume-based criteria prevent bright radio emission generated by AGN-driven shocks from being misclassified as relics, merger shocks may host embedded AGN, which can act as radio point sources and artificially boost the total luminosity.
To mitigate this issue, we iteratively remove the brightest gas cells in the radio relic until the removal of any single cell results in a change of less than $10\%$ in the total radio luminosity.

The physical properties of radio relics, including magnetic field strength, Mach number, and the total shock dissipated energy, are derived using the cells that belong to each relic group. 
The relic center and its clustercentric distance are defined using the radio-emissivity-weighted average position and its distance to the cluster center, respectively. 
We estimate the LLS using the maximum separation between relic cells with radio luminosity above $10^{19} \rm~W~Hz^{-1}$. 
Finally, we exclude the radio structures with $\rm LLS < 0.3 \rm~Mpc$ or a radio emissivity-weighted Mach number of $\mathcal{M} > 10$ to remove the radio emission generated by non-merger-driven sources. 

At the node of large-scale structures, clusters are rich in cosmological shockwaves \citep[e.g.,][]{2003ApJ...593..599R}, which can also produce faint radio emission. To avoid confusing these faint non-merger-shock features with relics, we exclude the radio features from the relic catalog when their luminosity is too low:
\begin{equation}
    L_{\rm 1.4GHz, 1}<5\times10^{20}\rm~W~Hz^{-1}~(M_{\rm 500c}/10^{14} M_{\odot})^{1.5}.
\end{equation}
This criterion is based on the mass–luminosity relation of simulated radio relics \citep{2024A&A...686A..55L}, where a non-detection is defined as cases in which the brightest radio feature is at least an order of magnitude fainter than predicted by the relation.
For the secondary radio relic, we apply a more relaxed threshold with a normalization of $5\times10^{19}\rm~W~Hz^{-1}$, which ensures the identification of double radio relic counterparts with a luminosity ratio as low as 0.1. 
We note that this criterion still identifies a radio relic that is an order of magnitude fainter than the faintest relic reported to date \citep[e.g.,][]{2024A&A...690A.222B}. 
The secondary radio relic is considered to be undetected if its flux falls below this criterion.
All identified radio relics are visually inspected for confirmation.

In general, projection effects can affect our analysis of radio relics. 
However, the projection effect is expected to be negligible for double radio relic systems, since such systems are most visible when the merger axis lies near the plane of the sky \citep[e.g.,][]{2018ApJ...862..160W, 2019ApJ...882...69G,2019MNRAS.489.3905S}.
Therefore, we utilize three-dimensional information in our analysis to derive scaling relations, assuming that the projection effect is minimal.
We compare the simulated relics with observed systems listed in Table~2 of \citet{2024A&A...686A..55L}, hereafter referred to as observations, which include all radio relics and radio relic candidates reported in previous studies.

%We compare the simulated relics with observations, which include both radio relics and radio relic candidates listed in Table 2 of \citet{2024A&A...686A..55L}.
%Except for \textsection\ref{sec:RRMH}, we use the two brightest radio features from the observed radio relics for the comparison, without accounting for the misaligned relics that may originate from different merger activity \citep[e.g.,][]{2024ApJ...966...38R}.

\begin{figure}
\centering
\includegraphics[width=\columnwidth]{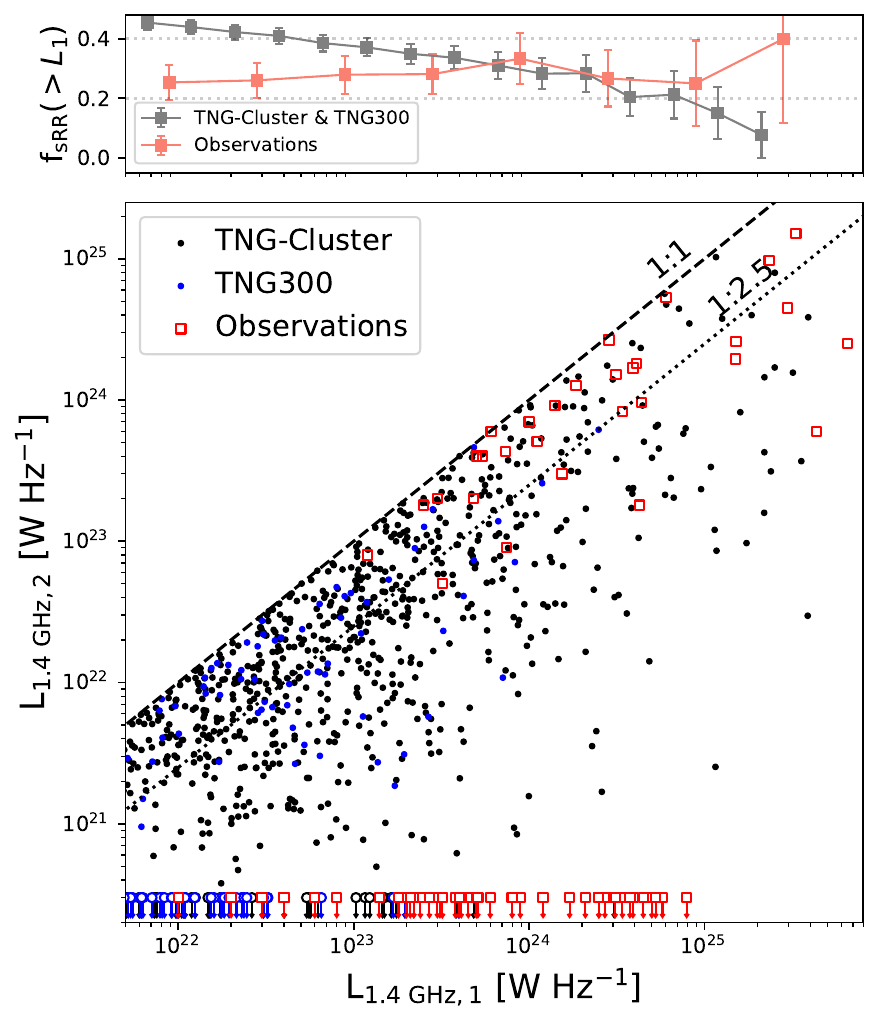}
 \caption{Radio luminosity at $1.4\rm~GHz$ of radio relics identified in the TNG-Cluster (black) and  TNG300 (blue) simulations, and in observations (red). The x- and y-axes show the luminosities of the primary and secondary relics, respectively. The luminosity of the secondary relic is marked with an upper limit when it is undetected. The dashed and dotted lines mark luminosity ratios of $1$ and $1/4$, respectively. The top panel shows the cumulative fraction of symmetric double relics with the primary relic luminosity above the given threshold, where symmetric double relic systems are defined as those with a luminosity ratio $>1/4$. The fraction of symmetric relics increases with a lower luminosity threshold.
 }
 \label{fig: lum_lum}
\end{figure}

\section{Statistics of double radio relics}
\label{sec:result}

\subsection{How rare are double radio relics?}
\label{sec: double_relics}

We identify 917 radio relic systems in total, consisting of 797 systems in the TNG-Cluster and 120 systems in the TNG300 simulations. Figure~\ref{fig: lum_lum} shows the 1.4~GHz radio luminosity of the primary and secondary relics in simulated systems and observations. 
%For systems where the secondary relic is not identified in simulations or remains undetected in observations, we assign an arbitrary upper limit in Figure~\ref{fig: lum_lum}. 
Although simulations are free from observational noise and projection effects, a substantial fraction of relic systems contain only a single detectable relic. % or feature a secondary relic with significantly lower luminosity.
Among the 917 systems, 122 host only one relic that satisfies our selection criteria, while 213 contain a faint secondary relic with a luminosity ratio below 0.1. These results indicate that even with deep observations, a significant fraction ($\gtrsim37\%$) of radio relic clusters may still appear to host only a single relic. We note that the number of asymmetric systems would be even larger, as some double relics in our sample may include cases where a single relic is misidentified as two due to fragmentation, or where single relics from separate mergers are mistakenly interpreted as a double relic system.

Both simulations and observations span a wide range of luminosity ratios between the primary and secondary relics in double radio relic systems. 
To quantify the number of double relics, we classify a system as hosting \textit{symmetric} double relics when the luminosity ratio between the secondary and primary relics exceeds 0.25. For example, this criterion identifies systems such as MACS J1752.0+4440 \citep[][]{2012MNRAS.425L..36V} and PSZ1 G096.89+24.17 \citep[][]{2014MNRAS.444.3130D} to host symmetric double relics.
The top panel of Figure~\ref{fig: lum_lum} shows the cumulative fraction of symmetric double relics as a function of the primary relic luminosity threshold, representing samples from radio surveys of varying depth. The fraction of symmetric systems is the lowest ($\mytilde20\%$) for clusters with bright primary relics ($L_{1.4~\rm GHz,1}>10^{25}\rm~W~Hz^{-1}$) and increases toward lower luminosity thresholds, reaching $\sim40\%$ for $L_{1.4~\rm GHz,1}>10^{22}\rm~W~Hz^{-1}$. This trend suggests that many clusters identified as hosting a single radio relic in shallow surveys may, in fact, reveal double relic systems when observed with deeper future surveys.

\begin{figure}
\centering
\includegraphics[width=0.9\columnwidth]{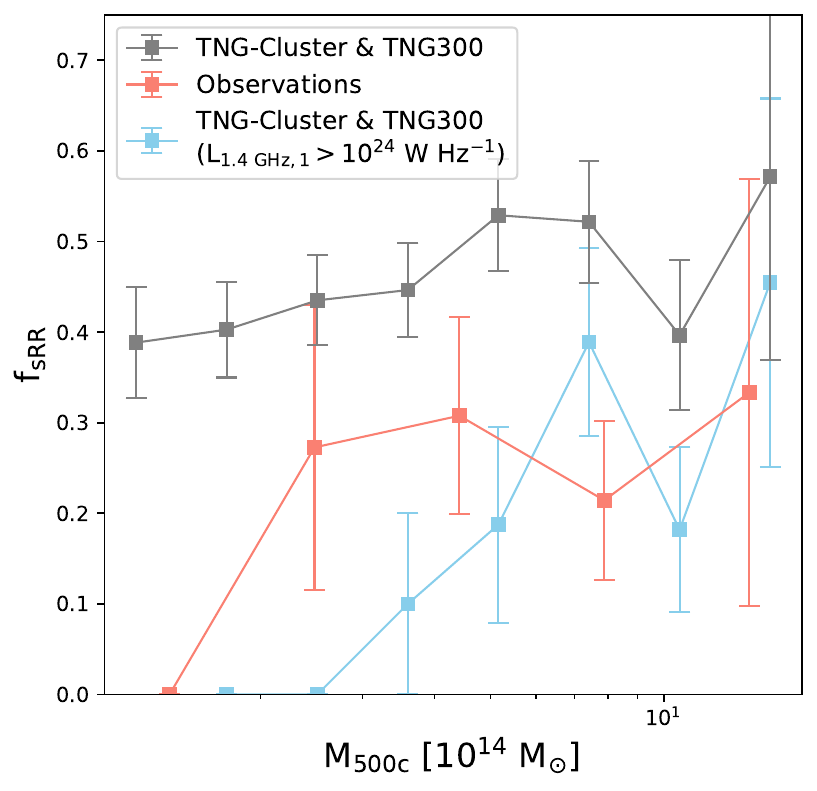}
 \caption{Fraction of symmetric double radio relics with $L_{\rm 1.4GHz,2}/L_{\rm 1.4GHz,1}>0.25$ in radio relic systems as a function of cluster mass. The black line shows the fraction from simulated radio relics in TNG-Cluster and TNG300, the red line represents observations, and the blue line shows the symmetric fraction of simulated relics with $L_{\rm 1.4GHz,1}>10^{24}\rm~W~Hz^{-1}$. The simulated radio relics yield a symmetric fraction comparable to that observed after applying the luminosity cut. 
 %Fraction of symmetric double radio relics with $L_{\rm 1.4GHz,2}/L_{\rm 1.4GHz,1}>0.25$ in radio relic systems (top) and full cluster samples (bottom) as a function of cluster mass. The black line shows the fraction from simulated radio relics in TNG-Cluster and TNG300, the red line represents observations. The blue line shows the symmetric fraction of simulated relics with $L_{\rm 1.4GHz,1}>10^{24}\rm~W~Hz^{-1}$. The simulated radio relics yield a symmetric fraction comparable to that observed after applying the luminosity cut. 
 }
 \label{fig: fdrr_mass}
\end{figure}

Observations, in contrast, show a relatively constant symmetric fraction ($\mytilde30\%$). 
This difference may reflect limitations of the assumed acceleration models or stem from selection effects present in both simulations and observations. For bright relic systems, the number of samples from both observations and simulations is limited as the brightest relics are associated with the most massive cluster mergers \citep[][]{2017MNRAS.470..240N,2024A&A...686A..55L}. For faint systems ($L_{\rm 1.4GHz,1}<10^{23}\rm~W~Hz^{-1}$), only a few examples have been reported from observations, likely due to sensitivity limits. When we select systems with $L_{\rm 1.4 GHz,1} > 10^{24}\rm~W~Hz^{-1}$ -- which assumes that a secondary relic would be detectable if $L_{\rm 1.4 GHz,2} > 2.5\times10^{23}\rm~W~Hz^{-1}$ -- both simulations and observations yield a comparable symmetric fraction of $\mytilde30\%$. We will further discuss the implications of the luminosity ratio in \textsection\ref{sec:eff}.  

Figure~\ref{fig: fdrr_mass} presents the symmetric fraction as a function of cluster mass. The fraction remains relatively uniform across the mass bins, with a marginal increase toward higher masses and an overall level of $40$–$50\%$. This weak trend indicates no strong dependence on cluster mass. The result differs from observations, which show a lower overall fraction and report only a small number of low-mass systems hosting double relics \citep[][]{2017MNRAS.472..940K,2017A&A...597A..15D,2024MNRAS.52710986C,2025ApJ...984...26A}. Since relics in low-mass clusters are intrinsically faint, the secondary relic may fall below detection thresholds or be misidentified due to projection effects. When applying a luminosity cut of $L_{\rm 1.4GHz, 1}>10^{24}\rm~W~Hz^{-1}$, the symmetric fraction becomes comparable to the observed values. The relatively higher fraction in the lowest mass bin from observation can be attributed to a few deep or low-frequency observations that detect faint double relics in low-mass cluster mergers \citep[e.g., Abell~1925;][]{2022A&A...660A..78B}.

Overall, double radio relics remain rare across the full cluster mass range. However, as the fraction increases significantly when fainter relics are included, the number of detected double relics will rise with the improved sensitivity of upcoming radio surveys capable of probing faint emission in low-mass cluster mergers. We will discuss the implications for future surveys in \textsection~\ref{sec: SKA}.

\begin{figure}
\centering
\includegraphics[width=0.9\columnwidth]{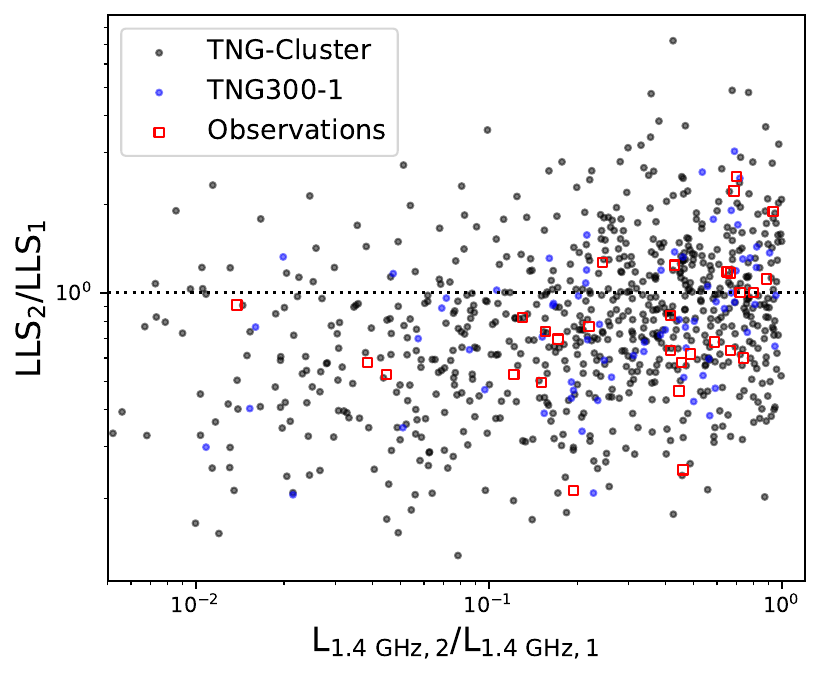}
 \caption{Ratio of the largest linear size between the primary and secondary relics as a function of their luminosity ratio in TNG-Cluster (black) and TNG300 (blue). Observed radio relics are shown as red points. While the overall size ratio broadly follows the luminosity ratio, many secondary relics exhibit a larger extent than their primary counterparts. 
 }
 \label{fig: size_n_lratio}
\end{figure}

\begin{figure*}
\centering
\includegraphics[width=2\columnwidth]{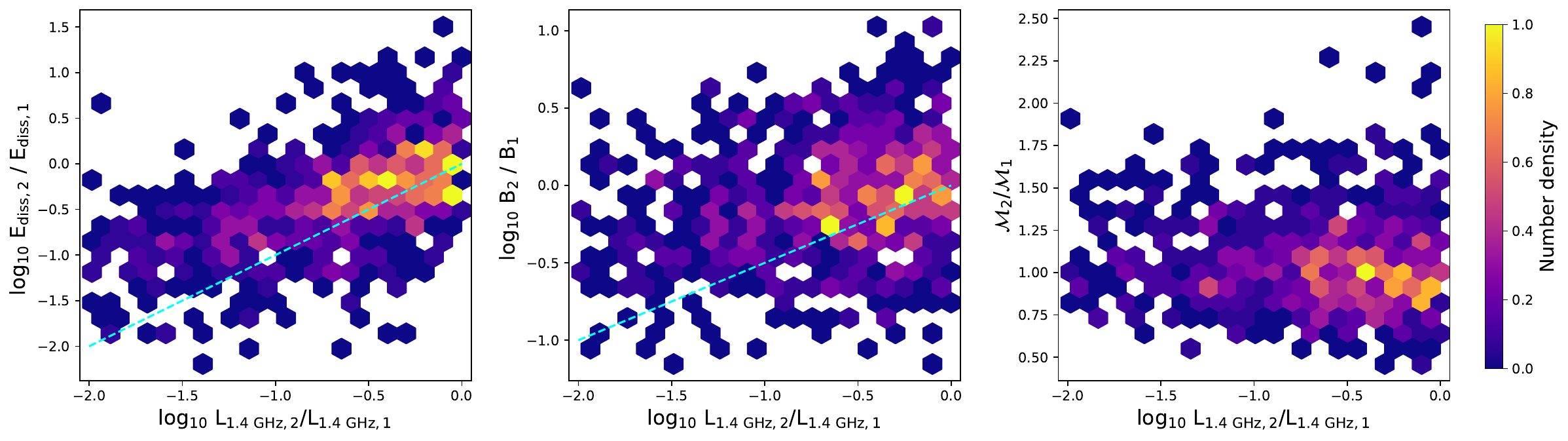}
 \caption{Ratio of the average shock properties between the primary and secondary relics compared to their luminosity ratio from TNG-Cluster and TNG300. From left to right, the net shock-dissipated energy ($E_{\rm diss,2}/E_{\rm diss,1}$), the mass-weighted average magnetic field strength ($B_{2}/B_{1}$), and the mass-weighted average Mach number ($\mathcal{M}_2/\mathcal{M}_1$) are shown as a function of the luminosity ratio. The color scale represents the number density of simulated systems in each bin. The cyan dashed line indicates the analytical relation between radio emissivity and the underlying cell properties, assuming a spectral index of $\alpha=-0.5$. The shock properties display broad variation, indicating that no single parameter alone determines the luminosity ratio.
 }
 \label{fig: relic_props}
\end{figure*}

\subsection{Properties of double radio relic pairs}
\label{sec: shock_properties}

Observed double radio relics exhibit a wide range of size ratios. While many bright systems show a larger extent in the primary relic \citep[i.e., the more luminous one; e.g., CIZA J2242.8+5301;][]{2010Sci...330..347V}, cases with a larger secondary relic have been reported in a few systems \citep[the less luminous relic is more extended; e.g., PSZ2 G277.93+12.34;][]{2024MNRAS.531.3357K}.
Figure~\ref{fig: size_n_lratio} compares the size and luminosity ratios between the primary and secondary radio relics from simulations and observations. 
In general, the size ratio decreases with decreasing luminosity ratio, since the primary relics tend to be larger than the secondary relics following the size-luminosity correlations. The Pearson correlation coefficient is 0.5.
Nevertheless, a substantial fraction ($\mytilde40\%$) of systems present a larger secondary relic, implying that factors beyond relative size contribute significantly to the luminosity ratio between the relics.

%The overall distribution of size and luminosity ratios in the simulations is broadly consistent with observations. 
The observed systems generally fall within the range spanned by the simulated size and luminosity ratios. 
Some systems lie near the edge of this distribution, notably those where re-acceleration of fossil CRe has been suggested, including ZwCl 1447.2+2619 \citep[$\rm LLS_2/LLS_1\sim0.25$, $\rm L_2/L_1\sim0.5$;][]{2022ApJ...924...18L} and ZwCl 0008.8+5215 \citep[$\rm LLS_2/LLS_1\sim0.21$, $\rm L_2/L_1\sim0.2$;][]{2011A&A...528A..38V}. %, and PLCK G287.0+32.9 \citep[$\rm LLS_2/LLS_1\sim0.6$, $\rm L_2/L_1\sim0.4$;][]{2014ApJ...785....1B}.
However, the simulated radio relics, which do not account for re-acceleration, also show ratios comparable to these observational examples.
This suggests that while fossil CRe may contribute to creating samples deviating from the distribution, not all radio relics with asymmetric relics require invoking fossil CRe.
Still, as many relics involving re-acceleration may only present a single detectable relic \citep[e.g., Abell 3411;][]{2017NatAs...1E...5V}, deeper observations are needed to verify the role of fossil CRe in the distribution of size and luminosity ratio.

To explore the origin of these differences, we examine the average property ratios of shock cells between the primary and secondary relics, as shown in Figure~\ref{fig: relic_props}. For shock-dissipated energy and magnetic field strength, we overlay correlations with slopes of 1 and 0.5, respectively, following the radio emissivity equation with the spectral index $\alpha=-0.5$ \citep[][]{2007MNRAS.375...77H}. 
The distributions broadly follow these trends but display substantial scatter, with some secondary relics exhibiting stronger shock dissipated energy or magnetic fields than the primary.
In particular, we find that $\mytilde22\%$ and $\mytilde38\%$ of the secondary relics exhibit higher shock dissipated energy or magnetic field strength than their brighter double relic counterparts, respectively.
These results suggest that the luminosity ratio of double relics is governed by a combination of shock parameters rather than a single factor.
The broad distribution of shock properties within relic cells may further contribute to the observed scatter, as small subregions with stronger Mach number or magnetic fields than the mass-weighted average can dominate the radio emission (e.g., \citealt[][]{2015ApJ...812...49H,2021MNRAS.506..396W}, see also Appendix \ref{sec: app_shock_property}). 
Together, these results imply that the diversity of observed luminosity ratios originates from variations in the internal properties of relics.

%The excess of secondary relics above the reference lines can also be attributed to the wide distribution of shock properties within relic cells, where the mass-weighted average may underestimate those of the subregions that dominate the radio emission \citep[e.g.,][]{2015ApJ...812...49H,2021MNRAS.506..396W}.
%Together, these results imply that the diversity of observed luminosity ratios originates from variations in the internal properties of relics.

\section{Merger history reconstruction with double radio relics}
\label{sec:RRMH}

%Radio relics (and their associated merger shocks) have been used as tracers of cluster merger history. Their elongated morphology, high polarization, and location provide information about the collision axis \citep[e.g.,][]{2011A&A...533A..35V,2019ApJ...882...69G}, the time since pericenter passage \citep[e.g.,][]{2021ApJ...918...72F}, and the mass ratio of colliding subclusters \citep[e.g.,][]{2011MNRAS.418..230V}. These interpretations are supported by idealized simulations that reproduce observed features through parameter tuning \citep[e.g.,][]{2007MNRAS.380..911S,2014ApJ...787..144L,2020ApJ...894...60L,2022MNRAS.509.1201C} and merging clusters identified in the cosmological simulations  \citep[e.g.,][]{2013ApJ...765...21S,2016MNRAS.459...70V,2018ApJ...857...26H,2025ApJ...978..122L}.

Radio relics have been widely used as tracers of cluster merger history, with their morphology, polarization, and position providing information on the merger geometry and phase.
Using statistical samples from TNG-Cluster and TNG300, we assess the reliability of merger history reconstruction based on radio relics. Galaxy clusters can undergo complex merger histories that generate multiple shock waves or modify shock properties through interactions with infalling subhalos \citep[e.g.,][]{2023ApJ...957L..16B,2024A&A...686A..55L}. In this analysis, we associate radio relics with the merger event involving the most massive sub-cluster, whose pericenter passage occurred between the current and the preceding full snapshots. If the same relic pair was already present in the preceding full snapshot, we retain the previously linked merger event. Merger history reconstruction generally uses double relic pairs located on opposite sides that form a parenthesis-like configuration, while relics located at asymmetric configurations are often excluded from such analyses \citep[e.g.,][]{2012A&A...546A.124V}. 
We address this by redefining the secondary relic in cases where two relics appear on the same side, replacing it with a third relic located on the opposite side to form a physically consistent pair.
These association and selection processes are verified through visual inspection, resulting in 578 double-relic systems and 845 systems in total, including single-relic systems.

\begin{figure}
\centering
\includegraphics[width=0.9\columnwidth]{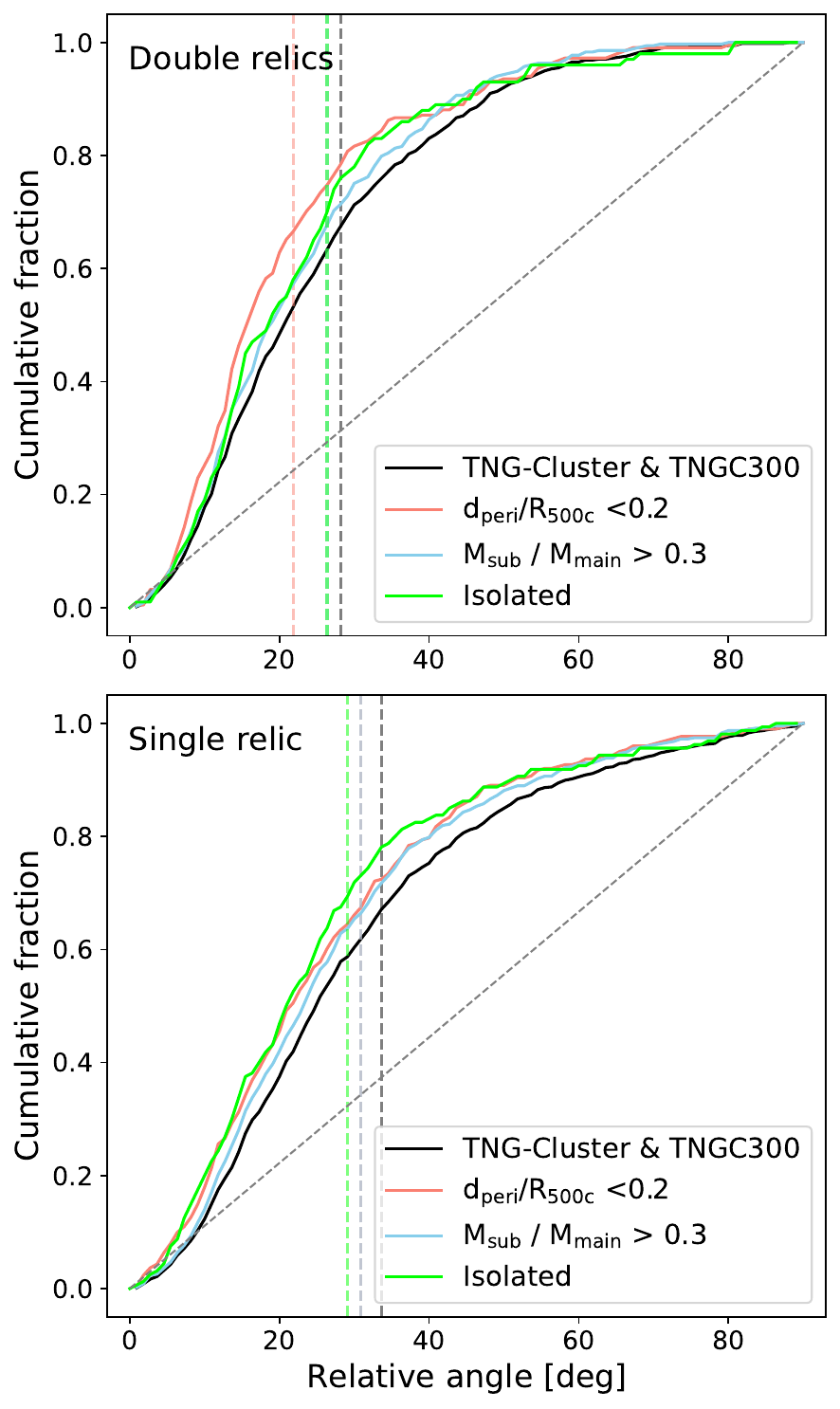}
\caption{
Cumulative fraction of the relative angle between the collision axis and the vector connecting the centers of the double radio relics (top) or the center of the primary relic (bottom) from TNG-Cluster and TNG300. Color show subsamples with simple merger geometries, selected as head-on mergers ($d_{\rm peri}/R_{\rm 500c} < 0.2$), major mergers ($M_{\rm sub}/M_{\rm main} > 0.3$), and isolated mergers. The results show that relic axes can typically constrain the collision axis within $\mytilde30^{\circ}$.
}
 \label{fig:collision_axis}
\end{figure}

\begin{figure}
\centering
\includegraphics[width=\columnwidth]{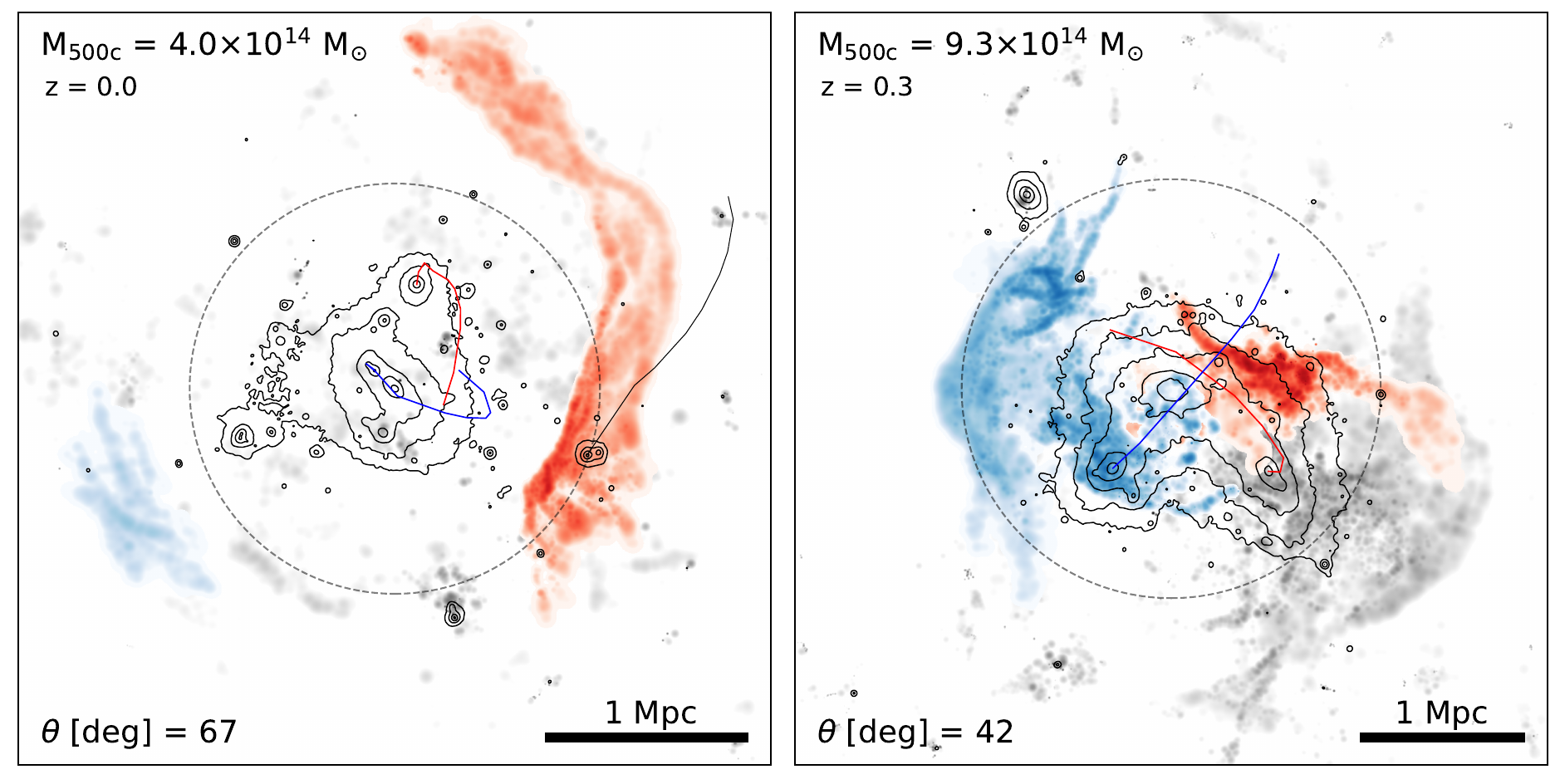}
 \caption{Examples of misalignment between the relic axis and the collision axis from TNG-Cluster and TNG300. The colormap shows the radio surface brightness, while the primary and the secondary radio relics are shown in red and blue colormaps, respectively. 
 The contours depict the dark matter surface density, and the dotted line shows the $R_{\rm 500c}$ of the cluster. 
 The red line marks the orbit of the most massive sub-cluster over the last $1.5\rm~Gyr$, while the blue line depicts the orbit of the second most massive sub-cluster during the same period. The black line on the left panel shows the orbit of the pre-merging halo that boosted the local surface brightness of the primary radio relics.}
 \label{fig:example_misaligned}
\end{figure}

\subsection{Relic axis and collision axis}
\label{sec:collision_axis}

The top panel of Figure~\ref{fig:collision_axis} shows the cumulative distribution of the angle between the collision axis and the line connecting the centers of double radio relics, hereafter referred to as the relic axis. In single relic systems, the relic axis is defined by the line connecting the radio relic center to the main cluster center.
For double relic systems, $68 \%$ of relic axes align with the collision axis within $28^\circ$, indicating that double relics effectively trace the merger geometry. The single relic can also constrain the merger geometry, while its alignment is weaker than double relics. As shown in the bottom panel, $68\%$ of the collision axes in the single relic systems lie within $35^\circ$ of the direction to the primary relic center. These results demonstrate that relics in general trace the merger axis, consistent with observations where $\mytilde80\%$ of relics align with the X-ray major axis within $30^\circ$ \citep{2011A&A...533A..35V}.

\begin{figure*}
\centering
\includegraphics[width=1.5\columnwidth]{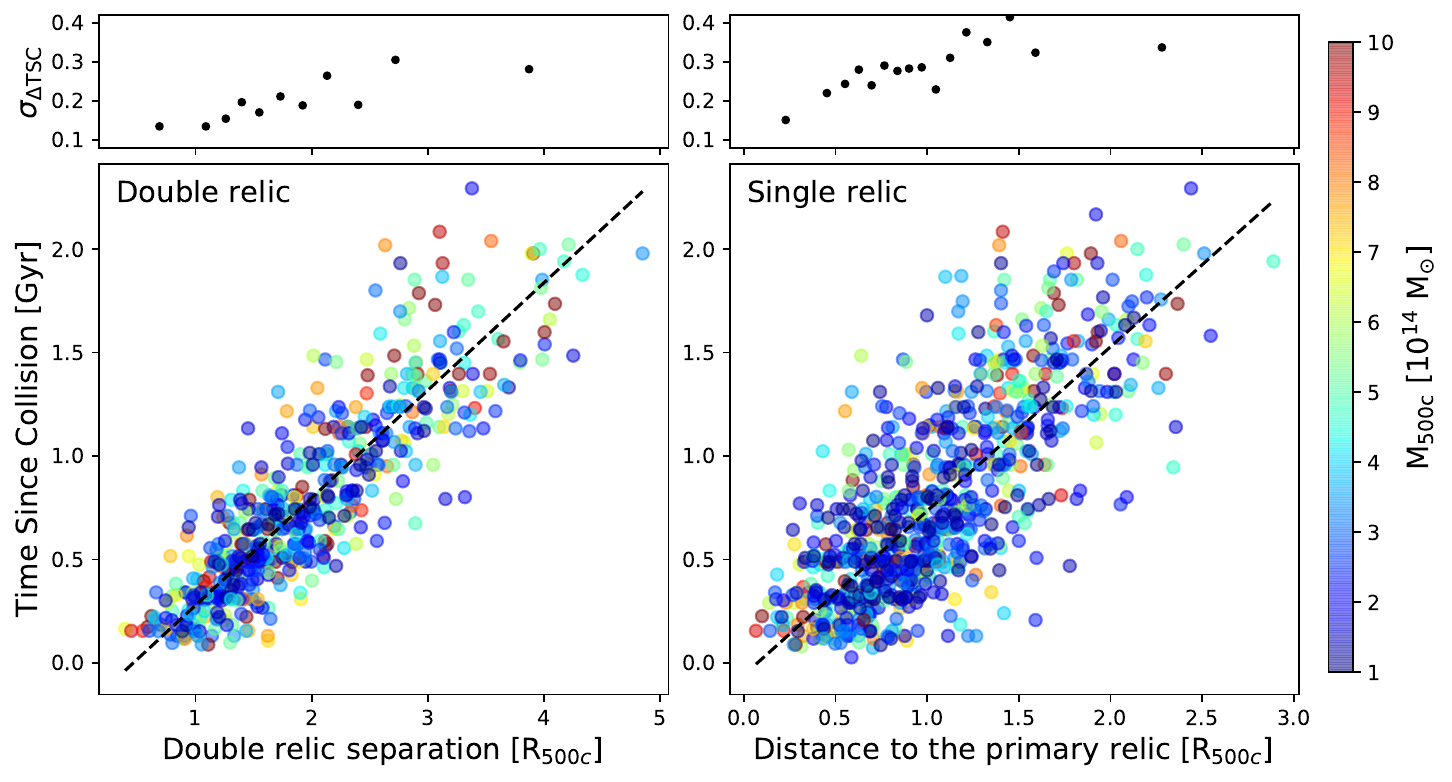}
 \caption{Correlation between the time since collision (TSC) and the double radio relic separation (left) or the single radio relic separation (right), both normalized by $R_{\rm 500c}$. Colors indicate $M_{\rm 500c}$ of the main cluster. The dashed lines represent the best-fit linear relations over the ranges $[1.0, 3.5]$ and $[0.5, 2.5]$ for double- and single-relic separations, respectively. The top panels present the standard deviation of the residuals from the relations. The scatter increases with relic separation, while the double- and single-relic separations constrain the time since collision to $\mytilde 0.2$ and $\mytilde 0.3~\mathrm{Gyr}$, respectively.}
 \label{fig:TSC_relic_sep}
\end{figure*}

To test whether alignment improves under simpler merger geometry, we examine head-on mergers ($d_{\rm peri}/R_{\rm 500c} < 0.2$), major mergers ($M_{\rm sub}/M_{\rm main} > 0.3$), and isolated mergers. 
Isolated mergers are identified based on the stellar mass of the third most massive galaxy within $2 R_{\rm 500c}$. This mass is measured within twice the stellar half-mass radius. If it is less than $10\%$ of the BCG's stellar mass, the system is classified as isolated.
%We define the system as an isolated merger when the stellar mass, measured within two times the stellar half-mass radius,  of the third most massive subhalo within $2 R_{\rm 500c}$ is $<0.1M_{\star,\rm BCG}$. 
As presented in Figure~\ref{fig:collision_axis}, all three subsamples show improved alignment. For head-on mergers, $68\%$ of the relic axes of double and single relic systems align with the collision axis within $24^\circ$ and $32^\circ$, respectively. In major mergers, the relic axes align within $27^\circ$ and $33^\circ$ for double and single relics, with isolated mergers following similar trends. 
Specifically, we highlight that the relic axis from single relics in isolated mergers can constrain the collision axis with the accuracy comparable to that of the double relic systems. 
This suggests that even single relics can be used to infer merger geometry if observations confirm the simple merger history of the host cluster \citep[e.g., Abell 115;][]{2019ApJ...874..143K, 2020ApJ...894...60L}.

On the other hand, large misalignments ($>60^\circ$) are found in $4\%$ of double relics and $10\%$ of single relic systems. These misalignments are driven by additional merger events that create substructures in radio relics. 
%While many subclusters fall into clusters along similar trajectories, some follow distinct paths, producing relics along different axes or altering the properties of those relics in those directions. 
While most subclusters fall in along the large-scale structure, some follow more complex trajectories, resulting in misaligned relics or altering relic properties along different axes.
Figure~\ref{fig:example_misaligned} presents two such examples. As denoted by the red and blue paths, both systems contain multiple subclusters that pass pericenter within a timeframe of $0.2–0.3~\rm Gyr$. In addition, the system in the left panel hosts another $6 \times 10^{12}\rm~ M_{\odot}$ subhalo that falls in and compresses the merger shock. This interaction enhances the surface brightness of the primary relic along its infall direction and shifts the relic center, resulting in a large misalignment of $67^\circ$. The right panel shows a triple merger, where two $\mytilde10^{14}\rm~M_{\odot}$ subclusters generate distinct relics along different axes. As a result, the secondary relic is identified as a composite of two merger shocks, producing a misalignment of $42^\circ$ relative to a single collision event. 

Although individual collision axes can depart from the relic axis, it is noteworthy that relics remain aligned with the associated collisions even in such complex systems. These examples demonstrate that relics with intricate morphologies can reflect a complex merger history while still preserving information about individual collision axes. 
It is therefore plausible that systems with complex relic configurations, such as Abell 2744 \citep{2011ApJ...728...27O,2021A&A...654A..41R} and Abell 746 \citep{2024ApJ...966...38R,2024ApJ...962..100H}, may likewise reflect multiple mergers, with individual relics encoding information about distinct components of the merger history. 

\begin{figure}
\centering
\includegraphics[width=\columnwidth]{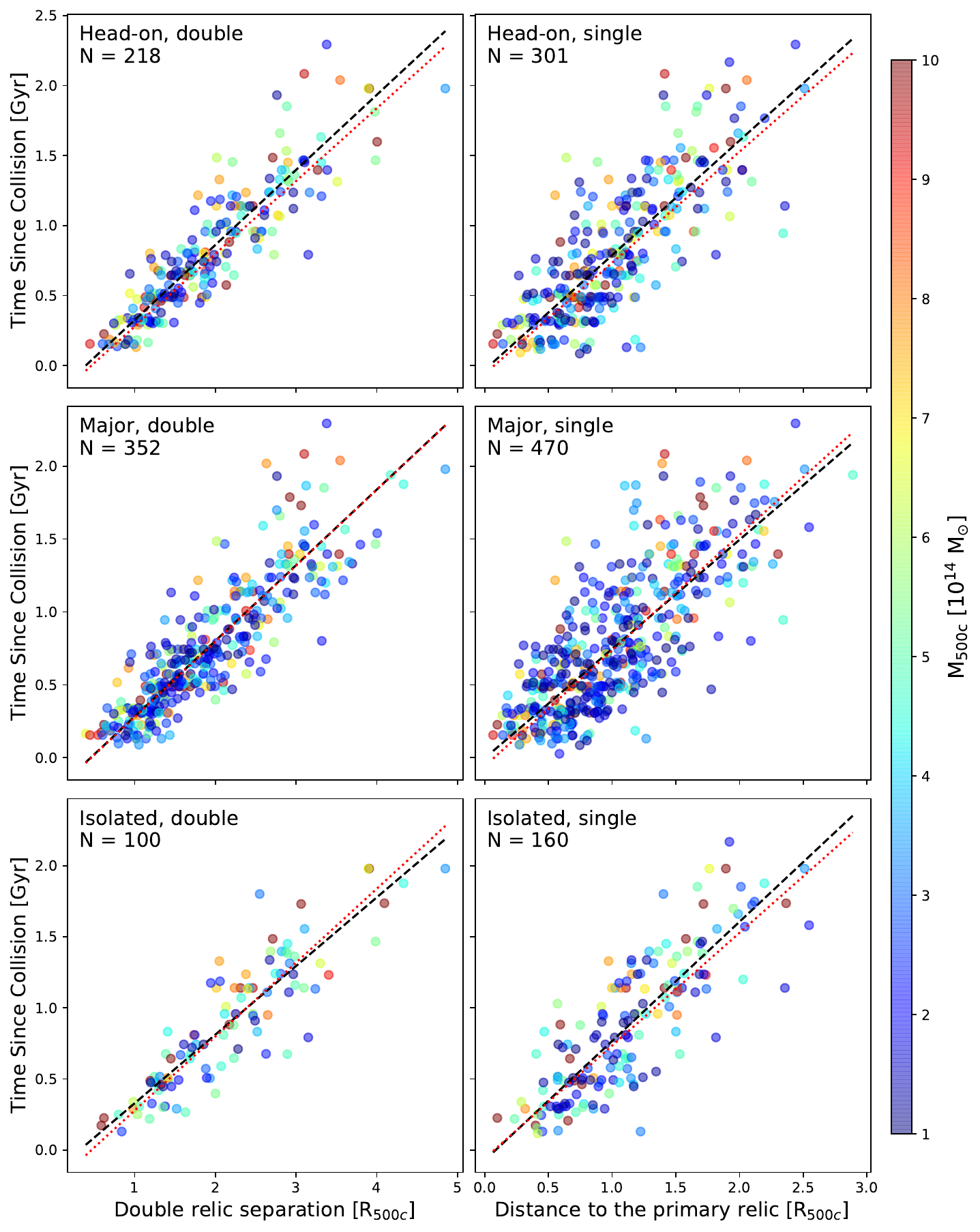}
 \caption{Same as Figure~\ref{fig:TSC_relic_sep}, but for subsamples with simple merger histories for double- (left) and single-relic systems (right). The subsamples are defined as in Figure~\ref{fig:collision_axis}. Linear relations are fitted over the same ranges as in Figure~\ref{fig:TSC_relic_sep}, with the red dotted line indicating the relation from the full samples. The subsamples show a comparable relation, with consistent or smaller variation from the relation. 
 }
 \label{fig:TSC_subsamples}
\end{figure}

\subsection{Relic Separation and the Time Since Collision}
\label{sec: TSC}

Unlike cluster components that remain bound to the gravitational potential, merger shocks continue to propagate outward after pericenter passage, making them effective probes of the TSC \citep[e.g.,][]{2023ApJ...945...71L}. Figure~\ref{fig:TSC_relic_sep} shows the correlations between TSC and the separation of double radio relics, hereafter denoted as $d_{\rm drr}$. For single relic systems, we use the distance from the BCG to the primary relic center ($d_{\rm srr}$ or single relic separation). 
We note that this definition of single relic separation can introduce scatter, since the primary relic, while generally expected to form ahead of the sub-cluster \citep[e.g.,][]{2018ApJ...857...26H}, may instead develop in front of the main cluster and remain in contact with it during the early merger phase \citep[e.g.,][]{2019MNRAS.488.5259Z}.
%, although primary relics are preferentially identified in front of the sub-cluster \citep[e.g.,][]{}, the single relics separation can be small at late merger phase when primary relic is generated by the main cluster merger shock and shock is yet detached from the cluster \citep[e.g.,][]{2019MNRAS.488.5259Z}. We note that this definition of single relic separation can create scatter, since the primary relic could be associated with the main cluster and in contact with the cluster at the early merger phase \citep[e.g.,][]{2019MNRAS.488.5259Z}.

Both double and single relic separations exhibit strong linear correlations with TSC. We fit the following relations:
\begin{equation}
    \rm{TSC} = \alpha  \frac{d_{\rm drr,srr}}{R_{\rm 500c}} + d_{o},
\end{equation}
where $\alpha$ is the slope and $d_{o}$ is the intercept. We perform linear regression over the ranges $d_{\rm drr}/R_{\rm 500c} = [1.0, 3.5]$ and $d_{\rm srr}/R_{\rm 500c} = [0.5, 2.5]$, excluding regimes where the TSC is incompletely sampled. The best-fit parameters are $(\alpha_{\rm drr}, d_{\rm o, drr}) = (0.52\pm0.02, -0.24\pm0.03)$ and $(\alpha_{\rm srr}, d_{ \rm o,srr}) = (0.79\pm0.03, -0.06\pm0.03)$ for double and single relic separation, respectively. The negative intercepts reflect that merger shocks are launched with a nonzero separation due to the ICM core and off-axis collision. The Pearson correlation coefficients are 0.83 and 0.73 for double and single relic separations, respectively, indicating a strong correlation between the TSC and relic separations.
We note that the correlation between relic and TSC is weaker when using the shock-inferred time, derived with the relic separation divided by the shock velocity (see Appendix \ref{sec: app_shock_velocity} for more detail).

The deviation of the true TSC of individual systems from the fitted relations can be interpreted as the uncertainty in the TSC estimate.
The top panel of Figure~\ref{fig:TSC_relic_sep} presents the standard deviation of the residuals between the fitted relation and the true TSC. For double relic systems, the uncertainty is $0.1–0.2\rm~Gyr$ for $d_{\rm drr}/R_{\rm 500c} < 2.0$ and increases to $\mytilde0.3\rm~Gyr$ at larger separations. For single relic systems, the uncertainty is larger than that of double relics, ranging from $0.2$–$0.3\rm~Gyr$ at $d_{\rm srr}/R_{\rm 500c} < 1.0$ and reaching $\mytilde0.4\rm~Gyr$ in later phases. The increasing uncertainty with time likely reflects variations in ICM profiles and cluster environments at large clustercentric distance. Still, since most observed double relic systems have $d_{\rm drr}/R_{\rm 500c} < 2.5$ \citep[][]{2025ApJ...984...24S}, we conclude that the TSC can be constrained with an accuracy of $\mytilde0.2\rm~Gyr$ using double relics. We highlight that this estimate relies solely on radio data, and the accuracy will improve by combining constraints from multi-wavelength observations.

The scatter in the relation likely arises from variations in merger geometry and dynamical complexity. Off-axis mergers with large pericenter distances can yield large relic separations even at early merger phases, while interactions with infalling subclusters can hinder shock propagation and reduce relic separations. Figure~\ref{fig:TSC_subsamples} shows the TSC-relic separation relations for the subsamples defined in \textsection\ref{sec:collision_axis}. Head-on and isolated mergers exhibit reduced uncertainties of $\mytilde0.17 ~\rm Gyr$ for double relics with $d_{\rm drr}/R_{\rm 500c} < 2.5$ and $\mytilde0.27 ~\rm Gyr$ for single relics with $d_{\rm srr}/R_{\rm 500c} < 1.5$, while major mergers present uncertainties comparable to those of the full sample. These results suggest that relic separations can constrain the TSC with the accuracy of $\lesssim0.2 ~\rm Gyr$ for double relics and $\lesssim0.3 ~\rm Gyr$ for single relics, when a simple merger geometry can be confirmed. We note that the TSC correlations derived from these subsamples are consistent with those obtained from the full sample.

We apply the relation to observed double radio relic systems using their WL mass estimates, or SZ-derived masses when WL data are unavailable. For El Gordo \citep[$d_{\rm drr}/R_{\rm 500c} \sim 1.2$;][]{2014ApJ...786...49L,2021ApJ...923..101K}, Abell 521 \citep[$\mytilde 1.4$;][]{2020ApJ...903..151Y,2024ApJ...962...40S}, CIZA J2242.8+5301 \citep[$\mytilde 1.7$;][]{2010Sci...330..347V,2015ApJ...802...46J}, ZwCl 1447.2+2619 \citep[$\mytilde 1.8$;][]{2022ApJ...924...18L}, Abell 3667 \citep[$\mytilde 2.6$;][]{2022A&A...659A.146D,2016A&A...594A..24P}, and PSZ2 G181.06+48.47 \citep[$\mytilde 2.9$;][]{2025ApJ...984...25R}, the estimated TSC ranges in $\sim[0.4,1.3]\rm~Gyr$, consistent with earlier studies that inferred TSC using X-ray morphology and idealized simulations \citep[e.g.,][]{2015ApJ...800...37M,2022ApJ...924...18L,2024A&A...689A.173O,2025ApJ...984...26A}. An exception is CIZA J2242.8+5301 \citep{2011MNRAS.418..230V}, where earlier modeling adopted a mass lower than the WL estimates. When we adopt that lower mass, our relation yields a TSC consistent with the previous study, highlighting the need for accurate mass measurements in the TSC estimate. In addition, the case of Abell 521 demonstrates how identifying both relics can further improve the accuracy of TSC estimation, where the uncertainty decreases from $\mytilde0.6 \pm 0.3 ~\rm Gyr$ when using a single relic to $\mytilde0.5 \pm 0.2 ~\rm Gyr$ when the newly detected counterpart is included\footnote{We note that the observed separation, measured between the shock front, differs from the definition of double relic separation used in Figure \ref{fig:TSC_relic_sep}, where adopted the emissivity-weighted centers. The projected separation measured between the shock edges can be $\mytilde15\%$ larger than the three-dimensional separation, corresponding to a $\mytilde0.1$–$0.2~\rm Gyr$ difference in the inferred TSC. We will further discuss the projection effect in \textsection\ref{sec: Projection}.}.

\begin{figure}
\centering
\includegraphics[width=\columnwidth]{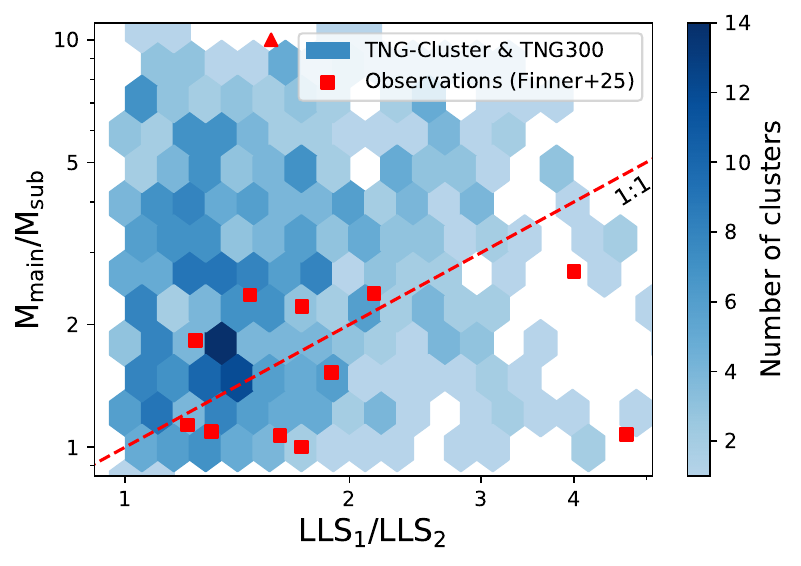}
 \caption{Size ratio of double radio relics versus the mass ratio of cluster mergers in TNG-Cluster and TNG300. Observational data from \citet{2025ApJS..277...28F}, based on the weak-lensing analysis, are shown as scatter points. PLCK G287.0+32.9 is marked with an upper limit due to its large mass ratio of $\mytilde {12} $. The dotted line indicates the one-to-one relation. While some observed and simulated systems show comparable size and mass ratios, many outliers are present in both observations and simulations.
 }
 \label{fig:size_mass_ratio}
\end{figure}

As discussed in \textsection\ref{sec:collision_axis}, radio relics produced by multiple mergers may retain information from individual collisions, so a single cluster can host relics that trace the TSC of different mergers. For instance, two prominent relics in Abell 2744 are located $1.3\rm~Mpc$ and $0.9\rm~Mpc$ from the cluster center in a similar direction \citep[][]{2017ApJ...845...81P}. Based on our relation, these correspond to sequential mergers that occurred $\mytilde0.8$ and $\mytilde0.5\rm~Gyr$ ago, which are consistent with estimates from idealized simulations \citep[e.g.,][]{2024arXiv240703142C}. In the Bullet Cluster, a relic candidate at $d_{\rm srr}/R_{\rm 500c} \sim 1.5$ at the northern outskirts \citep[][]{2023MNRAS.518.4595S} complements the well-known bullet shock fronts separated by $\mytilde1.5\rm~Mpc$ from the primary relic \citep[][]{2002ApJ...567L..27M,2015MNRAS.449.1486S}. This configuration suggests that the system may have undergone at least two mergers; an older event $\mytilde1.1\rm~Gyr$ ago followed by a more recent collision $\mytilde0.3\rm~Gyr$ ago. These scenarios are supported by the detection of substructure and mass bridges in recent gravitational lensing analyses (\citealt[][]{2024ApJ...961..186C,2025ApJ...987L..15C}; Cho et al.\ in prep). 
We note, however, that these relic candidates may have alternative origins unrelated to merger activity \citep[e.g.,][]{2022ApJ...933..218B}. If confirmed as relics, they would highlight the potential of radio relics to reveal complex merger histories in dynamically active clusters.

Even when the double relics are located along the same axis, the discrepancy in the time estimates may hint at multiple merger events. For example, PLCK G287.0+32.9 presents two radio relics separated by $\mytilde3.1\rm~Mpc$ \citep[$\mytilde2R_{\rm 500c}$;][]{2014ApJ...785....1B,2017ApJ...851...46F}. However, only the secondary relic lies at a large distance of $\mytilde2.8\rm~Mpc$ ($\mytilde1.8R_{\rm 500c}$), while the primary relic is close to the cluster center ($\mytilde0.3R_{\rm 500c}$). This yields a TSC of $\mytilde0.8\rm~Gyr$ when using the double relic separation, but $\mytilde0.2\rm~Gyr$ when using only the distance to the primary relic. These discrepancies suggest that the two relics may originate from distinct merger events, or that the primary relic formed during a more recent second pericenter passage while the secondary relic resulted from the first pericenter passage, as proposed by \citet[][]{2014ApJ...785....1B}. However, care is needed when applying this method, as merger shocks can remain near the cluster core before they are detached \citep[e.g.,][]{2019MNRAS.488.5259Z} and as primary relic might be associated with the main clusters(e.g., RX J0603.3+4214, \citealt{2016ApJ...817..179J}; RXC J1314.4-2515, \citealt{2025ApJS..277...28F}).

\begin{figure}
\centering
\includegraphics[width=\columnwidth]{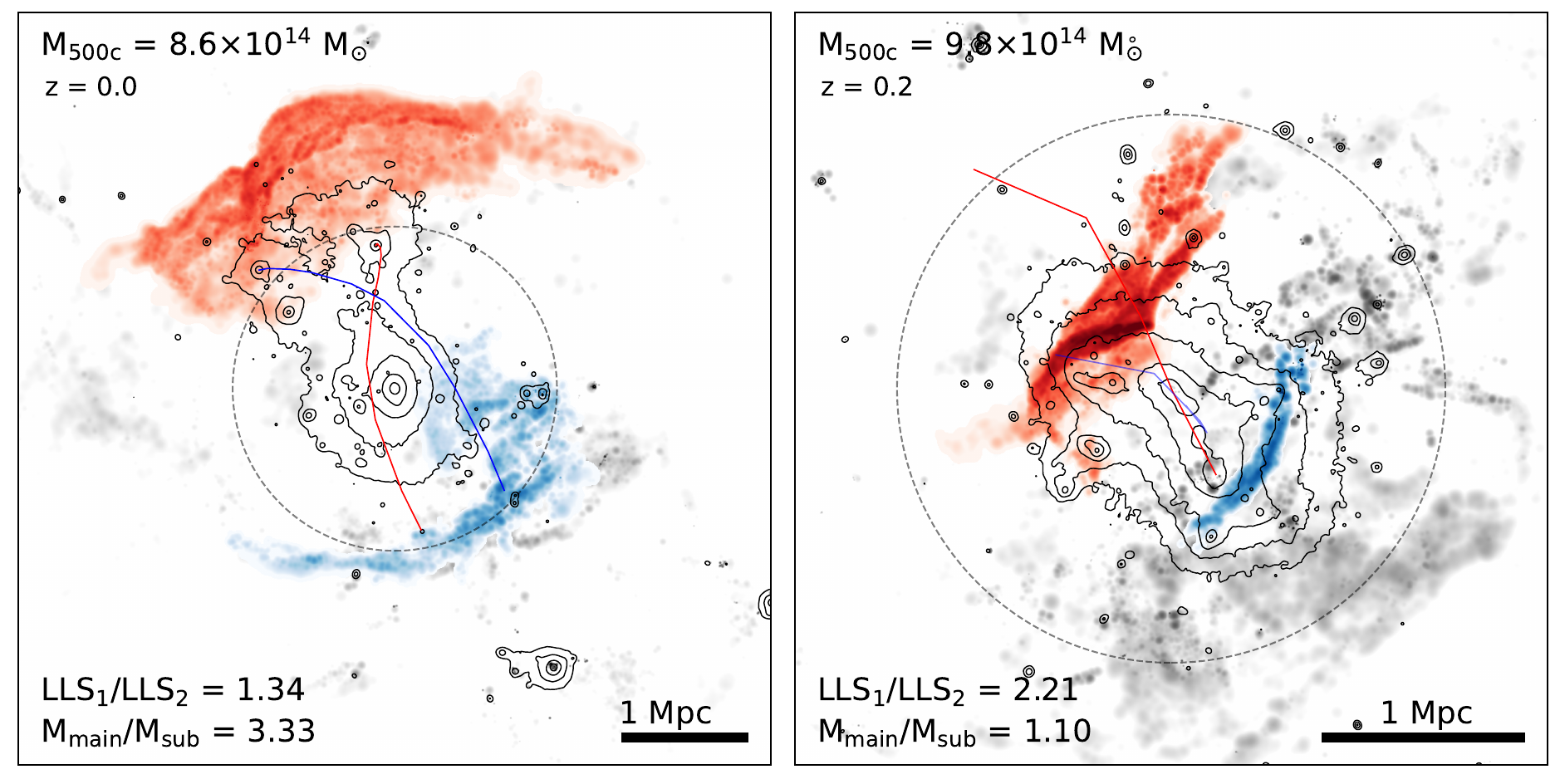}
 \caption{Two examples with different relic size and mass ratio from TNG-Cluster and TNG300. Relics formed during complex merger histories can exhibit a size ratio that does not follow the cluster mass ratio.
 }
 \label{fig:example_sizeratio}
\end{figure}

\subsection{Relic size and the cluster mass ratio}
\label{sec: Massratio}

Figure~\ref{fig:size_mass_ratio} shows the distribution of the size ratio of the primary and secondary relics with respect to the cluster mass ratio identified in TNG-Cluster and TNG300. 
No significant correlation is found between the two quantities, with a Pearson correlation coefficient of 0.03.
A large fraction of mergers with near-unity mass ratios produce relics of comparable size, with $82\%$ of systems with mass ratios smaller than 2 exhibiting relic size ratios below 2.
However, interestingly, we find that several minor mergers with mass ratios greater than 5 also display double relics of similar extent, with $68\%$ showing size ratios below 2. 
These results suggest that relic size ratio alone does not reliably trace the mass ratio of the colliding clusters.
We note that the Pearson correlation coefficient remains below 0.1 even in subsamples restricted to head-on or isolated mergers, indicating that the relic size ratio is determined by multiple parameters rather than by a few merger parameters alone.

This weak correlation is also evident in observations. The mass ratios of the observed data points in Figure~\ref{fig:size_mass_ratio} are constrained with WL analysis, which does not rely on hydrostatic assumptions \citep{2023ApJ...945...71L,2025ApJS..277...28F}. While some observed systems show good agreement between relic size and mass ratio (e.g., ZwCl 1447.2+2619, ZwCl 2341+0000), multiple systems exhibit large discrepancies. For example, ZwCl 0008.9+5215 is a major merger with nearly equal mass ($2.9$ and $2.7\times10^{14}~M_{\odot}$), yet its relic size ratio is close to 5 \citep[][]{2011A&A...528A..38V}. Similarly, MACS J1752.0+4440 is an equal-mass merger but hosts one of the brightest relic systems with a size ratio of about 2 \citep[][]{2012MNRAS.425L..36V}.

As with the collision axis and TSC, complex merger histories likely contribute to the scatter in the size-mass ratio relation. Figure~\ref{fig:example_sizeratio} presents two examples, one with a mass ratio of $\mytilde1:3$ and another with a near-equal mass ratio. Both systems show relic size ratios largely different from their mass ratios, which can be attributed to additional merger events occurring alongside the main cluster collision that produced the relics. The WL observations similarly report third substructures in systems with inconsistent size–mass ratios \citep[e.g., ZwCl 0008.9+5215, MACS J1752.0+4440;][]{2021ApJ...918...72F,2025ApJS..277...28F}. However, we note that some systems with substructures still exhibit consistent mass and size ratios. % (e.g., ZwCl 1447.2+2619; ZwCl 2341+0000).

Previous idealized \citep[e.g.,][]{2011MNRAS.418..230V,2022ApJ...924...18L} and cosmological simulations \citep[e.g.,][]{2025ApJ...978..122L} have shown that relic size ratios can follow the mass ratios of colliding systems. While statistical samples from TNG-Cluster and TNG300 include examples where size and mass ratios agree, many relics display inconsistent ratios. Based on these results, we conclude that relic size ratios cannot be used as a reliable proxy for cluster mass ratios. Still, as the measured size of radio relics strongly depends on observational depth \citep[e.g.,][]{2011ApJ...727L..25B,2021A&A...656A.154H}, a direct comparison with realistic mock observations is required for a more robust assessment of whether relic size can trace cluster mass ratios.

\section{Discussion}
\label{sec:discussion}

\subsection{Luminosity ratio statistics as a probe of plasma acceleration model}
\label{sec:eff}

\begin{figure}
\centering
\includegraphics[width=0.9\columnwidth]{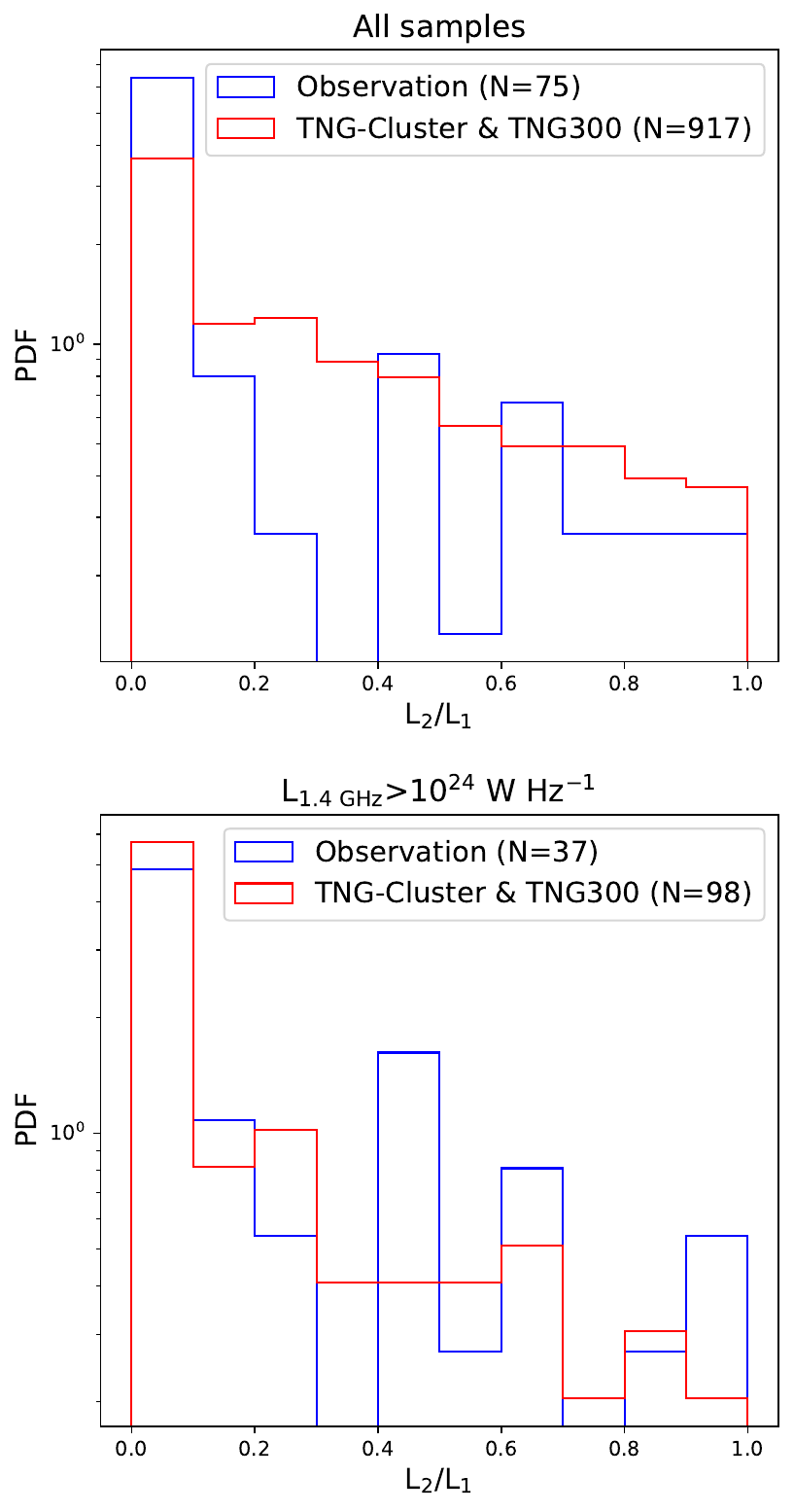}
 \caption{Radio luminosity ratio of the primary to secondary radio relics for all double relic samples (top) and the bright radio relic samples ($L_{1.4\rm~GHz, 1}>10^{24}\rm~W~Hz^{-1}$, bottom) from observation (blue) and the TNG-Cluster and TNG300 simulations (red). The single radio relic systems are included in the lowest bin. Both observation and simulation present a similar fraction of single radio relic systems ($\mytilde65\%$). 
 }
 \label{fig: lum_ratio}
\end{figure}

In \textsection\ref{sec: double_relics}, we estimate the luminosity ratios and compare the fraction of symmetric double radio relics in simulations and observations. Observational data show a relatively constant luminosity ratio across different cluster masses and radio powers, in contrast to simulations where the fraction of symmetric relics increases toward lower masses and fainter luminosities. However, when a luminosity cut is applied to the simulated sample, the resulting fractions become consistent with the observed values.

This agreement is notable given that our simulations assume a simple acceleration model applied uniformly to both merger shocks. Previous observational studies have proposed alternative plasma acceleration models, such as the re-acceleration of fossil CRe, to address the low acceleration efficiency in weak shocks and to explain the absence of double relic counterparts in some systems \citep[e.g.,][]{2016ApJ...823...13K}. Following this idea, we suggest that the statistical distribution of radio relic luminosity ratios can serve as a novel testbed for acceleration models. In particular, we consider the possibility that, if fossil plasma re-acceleration from AGN is the dominant mechanism behind observed relics, the resulting luminosity ratio distribution would be more asymmetric than that predicted by simulations.
Although some systems exhibit rapid redistribution of fossil plasma injected by AGN activity  \citep[e.g.,][]{2021ApJ...914...73Z,2022A&A...661A..92B,2023ApJ...957L...4L,2024Galax..12...19V}, we consider that the fossil plasma content—and thus the acceleration efficiency—may differ between the two relics in a given pair due to their large spatial separation across the cluster  \citep[][]{2023Galax..11...45V}. 

Figure~\ref{fig: lum_ratio} compares the distributions of double radio relic luminosity ratios in simulations and observations. The simulated relics show a continuous luminosity ratio distribution, with approximately $\mytilde40\%$ of systems having a luminosity ratio below 0.25, compared to $\mytilde60\%$ in observations. However, since observations are more sensitive to bright radio relics, the measured luminosity ratios are likely biased by the non-detection of faint secondary relics. To account for this, the bottom panel of Figure~\ref{fig: lum_ratio} shows the luminosity ratio distribution after applying a luminosity cut of $10^{24}~\rm W~Hz^{-1}$, assuming that secondary relics brighter than $10^{23}~\rm W~Hz^{-1}$ would be identified. With this luminosity threshold, the distributions from simulations and observations become comparable, with both showing that $\mytilde70\%$ of systems have luminosity ratios below 0.25. This suggests that in bright radio relic systems, asymmetric double relics are common when assuming identical DSA conditions for both merger shocks.

We also note that the simulated asymmetric fraction may still be underestimated. In this study, the secondary relic is defined as the second-brightest radio feature in the field. This approach may occasionally identify a secondary relic located at a different clustercentric distance, potentially originating from a different mechanism. Such cases can yield large luminosity ratios, but the selection of the second-brightest feature can suppress the measured asymmetry. Projection will also increase the fraction of asymmetric relics in the simulated sample (\textsection\ref{sec: Projection}).

This demonstrates how luminosity ratio statistics can be used to test plasma acceleration models. Here, we have deliberately selected the brightest radio relics given the varying depth of available observations, which, however, significantly reduced the number of samples. On the other hand, future radio surveys, such as the SKA, will provide a wide-field radio map with consistent observational depth, enabling statistical analysis of the relic luminosity ratio for testing acceleration models. 
We relied solely on radio relics in this analysis, while multi-wavelength observations can complement the analysis by independently constraining the merger geometry. 
Combined, we propose that statistics of double radio relic luminosity ratios could provide a new diagnostic for testing plasma acceleration models, in addition to detailed case studies of individual systems with well-characterized merger histories. 

\subsection{How many double radio relics will be detected?}
\label{sec: SKA}

\begin{figure}
\centering
\includegraphics[width=0.9\columnwidth]{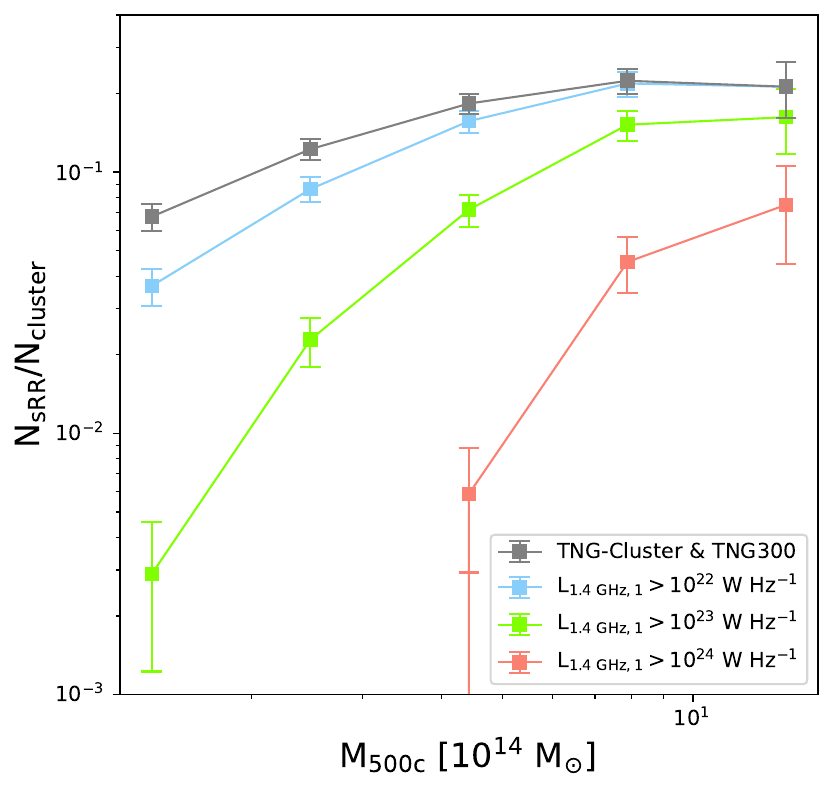}
 \caption{Fraction of clusters with symmetric radio relic compared to the total number of clusters in each mass bin, shown for varying luminosity cut from TNG-Cluster and TNG300. The luminosity criterion is applied to the primary radio relic, and the uncertainty is assumed to be Poisson noise. The dominant fraction of symmetric relics is found in the most massive clusters with a large luminosity cut, while the contribution from low-mass clusters rapidly increases with deeper surveys.
 }
 \label{fig: fdrr_mass_full}
\end{figure}

Although the simulation predicts a comparable fraction of asymmetric radio relics as observed, the fraction of double radio relics becomes substantially higher when no detection limit is applied. Detection of faint features has so far been possible only for a few clusters with deep targeted observations \citep[e.g.,][]{2022SciA....8.7623B,2024ApJ...962...40S}. However, as next-generation surveys, including LOFAR2.0 and SKA, will dramatically improve sensitivity, a large number of radio relic systems are expected to reveal double relic counterparts that remain undetected with current facilities. Here, we estimate the number of double radio relics expected to be detected with these upcoming surveys.

We adopt simplified conditions to obtain an order-of-magnitude estimate.
First, we consider an ideal configuration in which the double relics are well projected and free from classification confusion.
Then, we assume a radio relic to have uniform surface brightness extending over $1~\rm Mpc$ in length and $100~\rm kpc$ in width. 
With an 8$''$ beam from Briggs weighting in the SKA-Mid and SKA-Low AA* configurations, a 6-hour integration is expected to reach sensitivities of $\mytilde30~\rm nJy~arcsec^{-2}$ at $1.3~\rm GHz$ and $\mytilde150~\rm nJy~arcsec^{-2}$ at $300~\rm MHz$\footnote{\url{https://sensitivity-calculator.skao.int/}}.
These sensitivities correspond to $3\sigma$ detections of relics with $L_{1.4\rm~GHz}\sim10^{22.5}~\rm W~Hz^{-1}$ at $z\sim0.4$ or $L_{1.4\rm~GHz}\sim10^{22.0}~\rm W~Hz^{-1}$ at $z\sim0.1$. Under these conditions, we assume that the SKA will be able to identify double radio relic pairs with a symmetric luminosity ratio (i.e., luminosity ratio $>0.25$) up to $z=0.4$ when the primary relic has $L_{1.4\rm~GHz,1}>10^{23}~\rm W~Hz^{-1}$.

Figure~\ref{fig: fdrr_mass_full} shows the fraction of symmetric double relics with luminosity ratios larger than 0.25 relative to the total number of clusters in each mass bin.
The wide mass coverage of TNG-Cluster and TNG300 provides more than 100 clusters per bin, except for the most massive bin, which contains 28 clusters.
With a high luminosity cut of $L_{1.4\rm~GHz,1}>10^{24}~\rm W~Hz^{-1}$, symmetric double relics are most common in massive clusters with $M_{500\rm c}\sim10^{15}~M_\odot$, where $\mytilde8\%$ of systems are expected to host double relics.
In contrast, no double relics are expected in the low-mass regime, as the cluster mass limits the maximum achievable radio luminosity \citep{2024A&A...686A..55L}.

Assuming this fraction does not evolve significantly with redshift, we estimate the expected number of double radio relics up to $z=0.4$, where most observed systems lie.
We integrate the halo mass function from \citet{2016MNRAS.456.2361B} over both mass and redshift, weighted by the relic fraction as a function of cluster mass derived from Figure~\ref{fig: fdrr_mass_full}.
With the luminosity cut of $L_{1.4\rm~GHz,1}>10^{24}~\rm W~Hz^{-1}$, we estimate that $\mytilde100$ symmetric double relics would be detectable in a survey area of 20,000~deg$^2$ across the full mass range, with $\mytilde25\%$ originating from massive cluster mergers with $M_{\rm 500c}>8\times10^{14}\rm~M_{\odot}$ and only $\lesssim2\%$ from low-mass mergers with $M_{\rm 500c}<3\times10^{14}\rm~M_{\odot}$. We note that the predicted number of double relics exceeds the number of observed double relics, due to the ideal assumptions of double relic detection.

With next-generation facilities, the contribution from low-mass cluster mergers is expected to become increasingly significant.
As shown in Figure~\ref{fig: fdrr_mass_full}, the fraction of clusters hosting faint double relics rises to a few percent when including systems with $L_{1.4\rm~GHz,1}>10^{23}~\rm W~Hz^{-1}$, and the decline in occurrence rate with decreasing mass becomes noticeably shallower compared to the brightest samples.
Although the individual fraction remains low ($\mytilde0.1\%$), the large number of low-mass clusters substantially increases the total number of detectable systems.
Consequently, the number of double radio relics is expected to roughly double in the massive cluster regime, while low-mass clusters that previously showed no double relics begin to contribute appreciably.
Under these conditions, we predict $\mytilde2000$ double radio relic systems within the 20,000~deg$^2$ survey area, with $\mytilde5\%$ originating from massive clusters ($M_{\rm 500c}>8\times10^{14}\rm~M_{\odot}$) and nearly half ($\mytilde48\%$) from low-mass clusters with $M_{\rm 500c}<3\times10^{14}\rm~M_{\odot}$.

These estimates are based on the idealized detection assumptions, which can overpredict the number of double relics.
In addition, our simulated sample may overestimate the occurrence of radio relics in low-mass cluster mergers, as many simulated low-mass systems correspond to high-redshift progenitors of massive clusters, and simulations generally predict a higher fraction of high-redshift relics than observed \citep[e.g.,][]{2023A&A...680A..31J,2025A&A...695A.215D}. 
Nevertheless, given the large number of low-mass clusters and their non-negligible fraction of symmetric relics, upcoming radio surveys are expected to uncover a substantial new population of such systems, with low-mass mergers likely to play a key role in future studies involving double radio relics.

\subsection{Projection effects}
\label{sec: Projection}

In this work, we analyze the relative properties and spatial configurations of double radio relics using the three-dimensional distribution of gas cells. 
%Although double radio relic systems are commonly identified as plane-of-sky mergers \citep[e.g.,][]{2018ApJ...862..160W,2019ApJ...882...69G}, 
Projection effects can hinder the identification of secondary relics by significantly reducing the surface brightness of their radio emission.
Moreover, projection can change the apparent separation between radio relics, potentially introducing biases into the TSC estimates.

\begin{figure}
\centering
\includegraphics[width=\columnwidth]{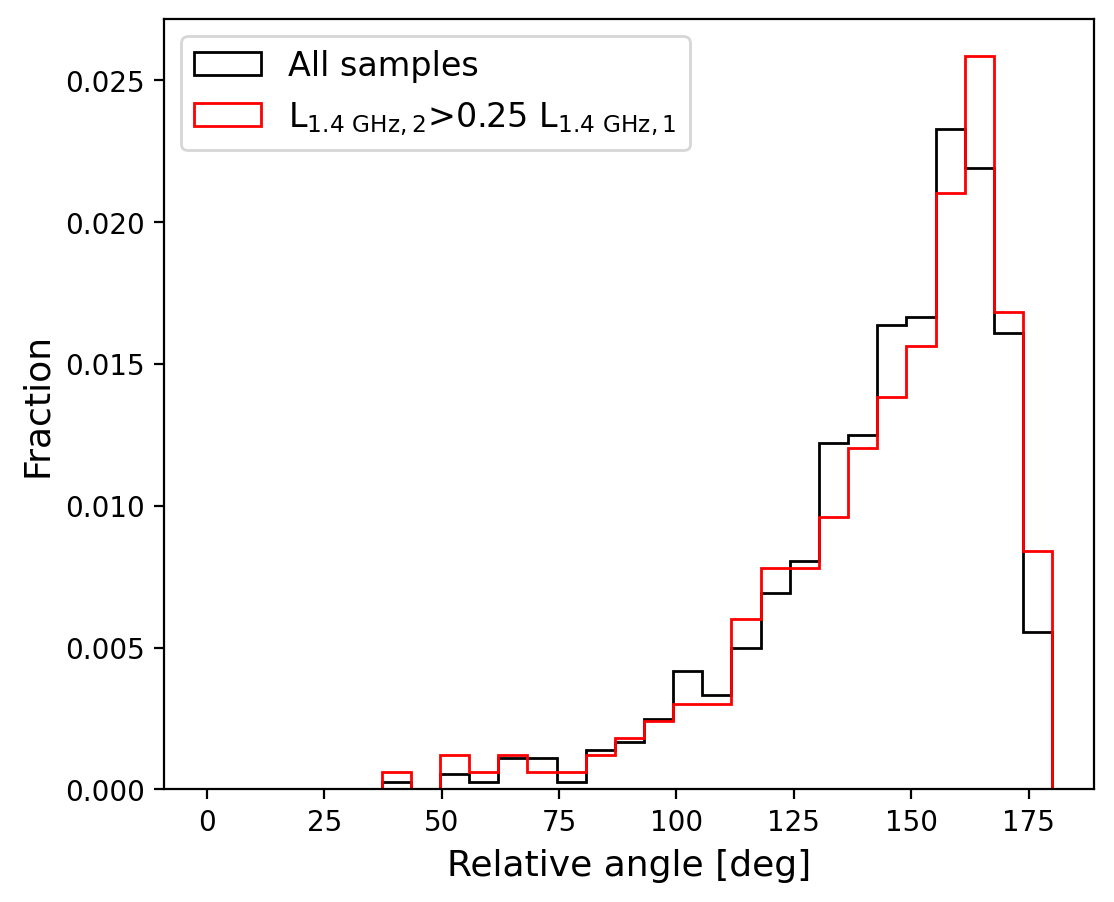}
 \caption{Probability distribution of the relative angle between the radio relic centers from TNG-Cluster and TNG300. The red histogram presents the distribution for the relics with $L_{1.4\rm GHZ, 2}/L_{1.4\rm GHZ, 1}>0.25$. The radio relic centers are relatively well-aligned, whereas a substantial fraction of radio relics can be misaligned. 
 }
 \label{fig: relic_rel_angle}
\end{figure}

We first analyze the potential non-detection of the double relics due to the projection effect. 
Figure~\ref{fig: relic_rel_angle} illustrates the distribution of the relative angle between the vectors to the centers of double radio relic pairs. The majority of relic pairs are located on opposite sides of the cluster, with relative angles exceeding $135^{\circ}$. However, $\mytilde25\%$ and $\mytilde10\%$ of relic pairs have relative angles greater than $45^{\circ}$ and $60^{\circ}$, respectively. The trend is similar for subsamples of radio relics with comparable luminosity ratios. 
This implies that even when primary relics are viewed edge-on and appear as textbook examples for radio relic cluster mergers, their secondary counterparts can be projected close to face-on, resulting in emission that resembles radio halos \citep[e.g.,][]{2013ApJ...765...21S,2015ApJ...812...49H,2024A&A...686A..55L}. As a result, the observed fraction of asymmetric radio relics may exceed that predicted by simulations.

\begin{figure}
\centering
\includegraphics[width=\columnwidth]{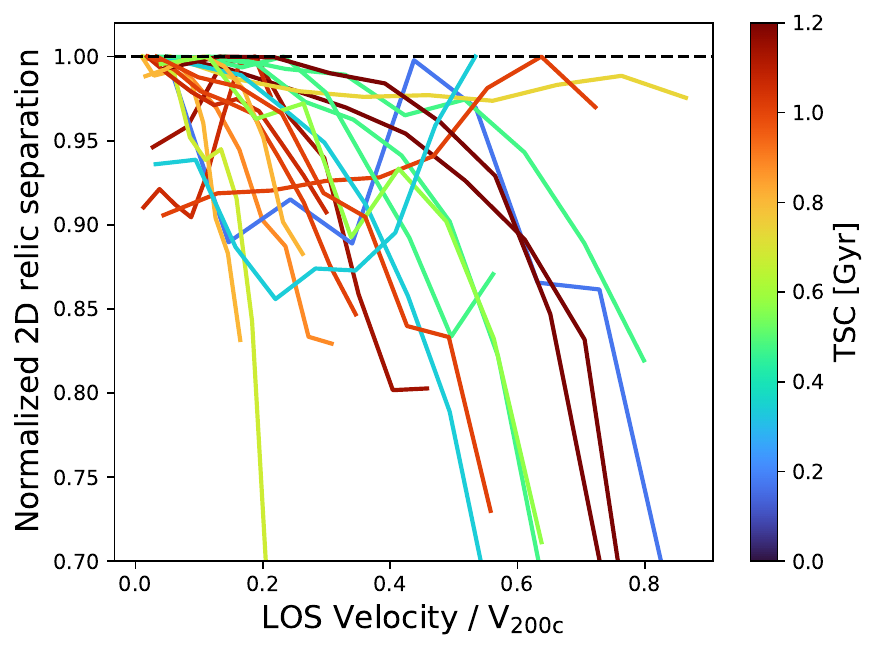}
 \caption{Double relic separation as a function of the relative line-of-sight velocity in 20 double relic systems at $z=0$ from TNG-Cluster. The separation represents the median value from 500 uniformly distributed projection maps and is normalized by the maximum separation. Colors indicate the TSC. The apparent relic separation changes only marginally with relative line-of-sight velocity, though the trend depends on the merger phase and relic morphology.
 }
 \label{fig:relic_sep_proj}
\end{figure}

We examine the impact of projection on relic separation, and consequently on the TSC estimate.
To minimize bias introduced by relic classification, we generate projected radio maps of the primary and secondary relics separately, using only the cells associated with each relic.
Our three-dimensional relic separation is measured using the radio-emissivity-weighted centers of the relics, which tend to lie closer to the cluster center than the shock edges. To follow the observational practice of using the shock front, we extract the radio surface-brightness profile from the projected radio map and compute the cumulative radio luminosity profile with the decreasing clustercentric distance. We identify the location of the shock front as the radius at which the cumulative flux profile reaches $10\%$ of the total radio flux. Due to the curved, shell-like morphology of relics, this procedure typically results in projected separations that are $\mytilde15\%$ larger than the corresponding three-dimensional emissivity-weighted separations.
We quantify projection using the relative line-of-sight (LOS) velocity of the merging clusters, measured from the subhalo velocities of the BCG analogs. We repeat the measurement for 500 uniformly distributed viewing directions, allowing us to characterize how separation changes with viewing angle with respect to the observable quantity.

Figure~\ref{fig:relic_sep_proj} presents the median double relic separations, normalized by their maximum separation, across different viewing angles in 20 double relic systems. In most cases, relic separation decreases with increasing LOS velocity, consistent with the expectation that larger viewing angles shorten the projected relic separation while increasing the relative LOS velocity. However, the projection effect on the relic separation is small, where the separation changes by $\mytilde10\%$ at $v_{\rm LOS} \lesssim 0.4 V_{\rm 200c}$ in many systems. For a $10^{15}\rm~M_{\odot}$ cluster merger, this corresponds to the relative LOS velocity of $\mytilde500\rm~km~ s^{-1}$, and so the projection effect on the relic separations would be negligible in CIZA J2242.8+5301 and MACS J1752.0+4440  \citep[][]{2019ApJS..240...39G,2019ApJ...882...69G}.

The projection effect depends on the merger phase. In early outgoing or late returning phases, where the relative velocity of the clusters is high, separations remain stable until $v_{\rm LOS} \gtrsim 0.5~V_{\rm 200c}$. On the other hand, when the cluster is near its apocenter passage, the relative velocity is low, and so the apparent relic separation can rapidly change even with a small relative LOS velocity. %This sensitivity underscores the importance of understanding the merger phase when applying TSC relations.
Moreover, the morphology of the radio relics also plays a role in the projection effect. For spherical shell-like relics, the relic separation only shows a small variation with different viewing angles, as the projection does not greatly change the distance to the edge of the spherical shell. In contrast, if the radio relic is linearly extended or presents irregular morphology, the apparent relic separation can quickly vary with the viewing angle, exhibiting flat or even inverted trends as in Figure~\ref{fig:relic_sep_proj}.

Overall, our results suggest that for systems with $v_{\rm LOS} \lesssim 0.4 V_{\rm 200c}$, projection effects introduce only a $\mytilde10\%$ uncertainty in separation, corresponding to an uncertainty of $\sim0.1$ Gyr in TSC. For mergers likely to be in the plane of the sky, this uncertainty translates into a small ($\sim0.1$ Gyr) underestimation in TSC using our relations in \textsection\ref{sec: TSC}. These degeneracies between LOS velocity, morphology, and separation further highlight the role of independent probes, including multi-wavelength data or machine-learning models that can model multiple physical parameters simultaneously \citep[e.g.,][]{2024ApJ...968...74S,2024arXiv241022416C}. 

\section{Summary} 
\label{sec:summary}

Double radio relics have been investigated through detailed analyses of individual systems, highlighting their distinctive morphologies and the complexity of cluster environments. With the increasing number of double relics detected in recent and upcoming radio surveys, statistical studies of double radio relics are becoming feasible. In this work, we utilize the cosmological MHD zoom-in simulation TNG-Cluster, combined with TNG300 of the IllustrisTNG project, to explore what can be inferred from the statistical properties of double radio relics. Our main findings can be summarized as follows:

\begin{itemize}
    \item The luminosity fraction of double relics shows wide variations, including many single-relic or asymmetric radio relic systems. This implies that a large number of clusters will only present a single radio relic even with deep radio observations due to their intrinsic asymmetry. The fraction of symmetric double relics increases toward lower luminosity and cluster mass, and matches observations once a luminosity threshold is applied (e.g., $L_{1.4\rm~GHz,1}\gtrsim10^{24}\rm~W~Hz^{-1}$). Massive clusters with $M_{\rm 500c} \sim 10^{15}\rm~M_{\odot}$ are the most likely hosts of these bright double relics.

    \item Double relics exhibit diverse shock properties. Primary relics are generally larger, but many systems contain larger secondary relics. The shock-dissipated energy and magnetic field strength also vary widely, with $\mytilde22\%$ and $\mytilde38\%$ of secondary relics exceeding their primary counterparts, respectively. These findings indicate that the luminosity asymmetry arises from the combined effects of multiple shock parameters rather than a single factor.

    \item The axis connecting double relics is generally aligned with the collision axis of the merging cluster. In the full sample, $68\%$ of systems have relic axes within $28^{\circ}$ of the collision axis, and the alignment is tighter in mergers with simpler geometries. Even in single-relic systems, the relic center can predict the collision axis within $\mytilde30^{\circ}$.
    
    \item The relic separation ($d_{\rm rr}$) correlates strongly with the time since collision (TSC). For double radio relic clusters with $d_{\rm drr}/R_{\rm 500c} =[1.0, 3.5]$, the best-fitting relation $\rm{TSC} = 0.52d_{\rm drr}/R_{\rm 500c}-0.24$ yield TSC estimates with an accuracy of $\mytilde0.2~\rm Gyr$. The accuracy is lower when using single relic separations, for which the best-fitting relation $\rm{TSC} = 0.79d_{\rm srr}/R_{\rm 500c}-0.06$ over $d_{\rm srr}/R_{\rm 500c} =[0.5, 2.5]$ gives an uncertainty of $\mytilde0.3~\rm Gyr$. The TSC estimates derived with these relations for observed radio relic cluster mergers are consistent with those inferred from X-ray observations and idealized simulations.
        
    \item Relic size ratios show only a weak dependence on the cluster mass ratio. Equal-mass mergers can produce highly asymmetric relics, while minor mergers may generate relics of comparable size.
    
    \item Complex merger histories are responsible for much of the variation between relic properties and merger parameters. Nevertheless, relics preserve signatures of past collisions, with individual shocks tracing separate events. This interpretation may hint at the complex merger history of the clusters hosting relics along different axes (e.g., Abell 746, the Bullet Cluster) or at different radii, yielding inconsistent TSC estimates (e.g., Abell 2744, PLCK G287.0+32.9).

    \item The distribution of double relic luminosity ratios provides a potential test for acceleration models. Both simulations and observations show that $\mytilde70\%$ of systems remain asymmetric after applying a luminosity cut ($>10^{24}\rm~W~Hz^{-1}$).
    This level of asymmetry is naturally reproduced by our model without invoking fossil CRe re-acceleration, indicating that such additional processes are not required to explain the observed asymmetry. Future wide-field radio surveys with larger samples will enable more stringent tests of these models.    

    \item Based on simple assumptions, future surveys such as SKA are expected to detect thousands of double relic systems. The majority are predicted to arise from low-mass cluster mergers, which hence will play a key role in future studies of cluster mergers using double radio relics.
    
    \item Projection effects can hinder the identification of secondary relics or alter apparent separations. In most cases, projected and three-dimensional double relic separations agree within $\mytilde10\%$ when the relative LOS velocity is below $0.3V_{\rm 200c}$ near pericenter or $0.1V_{\rm 200c}$ near apocenter.

\end{itemize}

The combination of TNG-Cluster and TNG300 provides a statistical framework for studying double radio relics across a wide range of cluster masses, offering guidance for interpreting the relic populations that will be revealed by upcoming surveys such as the SKA. In this work, we focused on summary statistics of their relative properties and their connection to merger histories, while relic morphology may also encode additional information on past collisions \citep[][]{2023ApJ...957L..16B, 2024A&A...686A..55L}. In this context, machine-learning approaches offer a promising way to extract further information from morphological complexity \citep[e.g.,][]{2024arXiv241022416C}. Together with realistic modeling of observational noise, machine-learning models trained on large simulated samples of relics could be applied to interpret next-generation survey data.

Our analysis focuses on properties inferred from radio relics, which are subject to degeneracies between parameters such as relic separation, where both the viewing angle and the TSC can affect the apparent distance. Previous studies have shown that multi-wavelength observations are capable of breaking such degeneracies by providing independent tracers of merger geometry and dynamics \citep[e.g.,][]{2019ApJ...882...69G,2024ApJ...968...74S}. A comprehensive view that combines radio and multi-wavelength data will be necessary for robust merger history reconstruction, and the ability of TNG-Cluster and TNG300 to produce mock observations across these bands offers a valuable reference for interpreting merger histories and plasma acceleration in upcoming surveys.

\begin{acknowledgements}
This work was supported by the National Research Foundation of Korea(NRF) grant funded by the Korea government(MSIT) (RS-2024-00340949).
M. J. Jee acknowledges support for the current research from the NRF of Korea under the programs 2022R1A2C1003130 and RS-2023-00219959.
D. Nelson acknowledges funding from the Deutsche Forschungsgemeinschaft (DFG) through an Emmy Noether Research Group (grant number NE 2441/1-1).
JAZ is funded by the Chandra X-ray Center, operated by the Smithsonian Astrophysical Observatory for and on behalf of NASA under contract NAS8-03060.
D. Nagai is supported by NSF (AST-2206055 \& 2307280) and NASA (80NSSC22K0821 \& TM3-24007X) grants. This work is also co-funded by the European Union (ERC, COSMIC-KEY, 101087822, PI: Pillepich).

The TNG-Cluster simulation suite has been executed on several machines: with compute time awarded under the TNG-Cluster project on the HoreKa supercomputer, funded by the Ministry of Science, Research and the Arts Baden-Württemberg and by the Federal Ministry of Education and Research; the bwForCluster Helix supercomputer, supported by the state of Baden-Württemberg through bwHPC and the German Research Foundation (DFG) through grant INST 35/1597-1 FUGG; the Vera cluster of the Max Planck Institute for Astronomy (MPIA), as well as the Cobra and Raven clusters, all three operated by the Max Planck Computational Data Facility (MPCDF); and the BinAC cluster, supported by the High Performance and Cloud Computing Group at the Zentrum für Datenverarbeitung of the University of Tübingen, the state of Baden-Württemberg through bwHPC and the German Research Foundation (DFG) through grant no INST 37/935-1 FUGG. 

This analysis was partially carried out on the VERA supercomputer of the Max Planck Institute for Astronomy (MPIA), operated by the Max Planck Computational Data Facility (MPCDF).

%\section*{Data Availability}
Both IllustrisTNG and TNG-Cluster simulations are publicly available and accessible at \url{www.tng-project.org/data}, as described in \cite{2019ComAC...6....2N}.
\end{acknowledgements}

\software{astropy \citep{2013A&A...558A..33A,2018AJ....156..123A,2022ApJ...935..167A}, Colossus \citep[][]{2018ApJS..239...35D}
          }

\appendix

\section{Mass-weighted and radio-weighted shock properties}
\label{sec: app_shock_property}

\begin{figure}
\centering
\includegraphics[width=0.85\columnwidth]{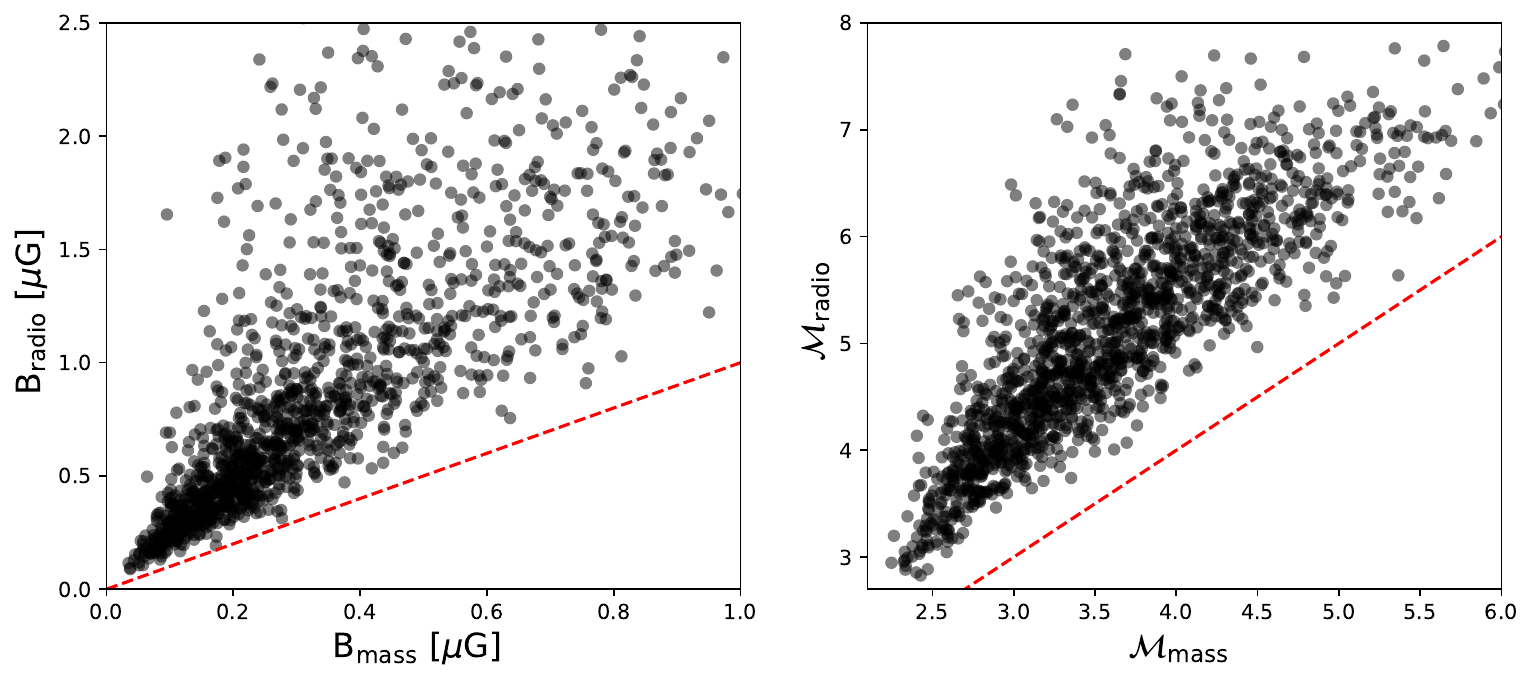}
 \caption{Comparison between the mass-weighted (x-axis) and radio emissivity-weighted (y-axis) averages of the magnetic field strength and Mach number of gas cells within radio relics.
The radio emissivity-weighted averages are systematically higher, reflecting the stronger sensitivity of emissivity to regions with high magnetic field strength and Mach number.
 }
 \label{fig: app_mass_radio_weight}
\end{figure}

Figure~\ref{fig: app_mass_radio_weight} compares the mass-weighted and radio emissivity-weighted average shock properties of gas cells within radio relics.
Both quantities show a clear correlation, with the radio emissivity-weighted averages systematically higher than the mass-weighted averages.
This difference arises because radio emissivity depends sensitively on both the magnetic field strength and Mach number, and is therefore weighted toward cells with stronger shocks and higher magnetic fields.
As noted in previous simulation studies \citep[e.g.,][]{2015ApJ...812...49H,2021MNRAS.506..396W,2025ApJ...978..122L}, the emissivity-weighted averages trace the upper end of the intrinsic property distributions, while the median values remain comparable to those derived from the mass-weighted averages.

\section{Time since collision estimate with relic separation and shock velocity}
\label{sec: app_shock_velocity}

Few observational studies have estimated the TSC by dividing the shock separation by the shock velocity \citep[e.g.,][]{2021ApJ...918...72F}. Following this approach, we compare the true TSC from the simulations with values obtained from the relic separation divided by the shock velocity. We estimate the shock velocity from the median Mach number of the relic cells and the sound speed derived using the center-excised temperature from the mass–temperature relation \citep{2010MNRAS.406.1773M}. For double radio relics, we adopt the average Mach number of the two relics. For brevity, we refer to the time inferred from double- and single-relic separations divided by the shock velocity as the shock-inferred time.

\begin{figure}
\centering
\includegraphics[width=0.85\columnwidth]{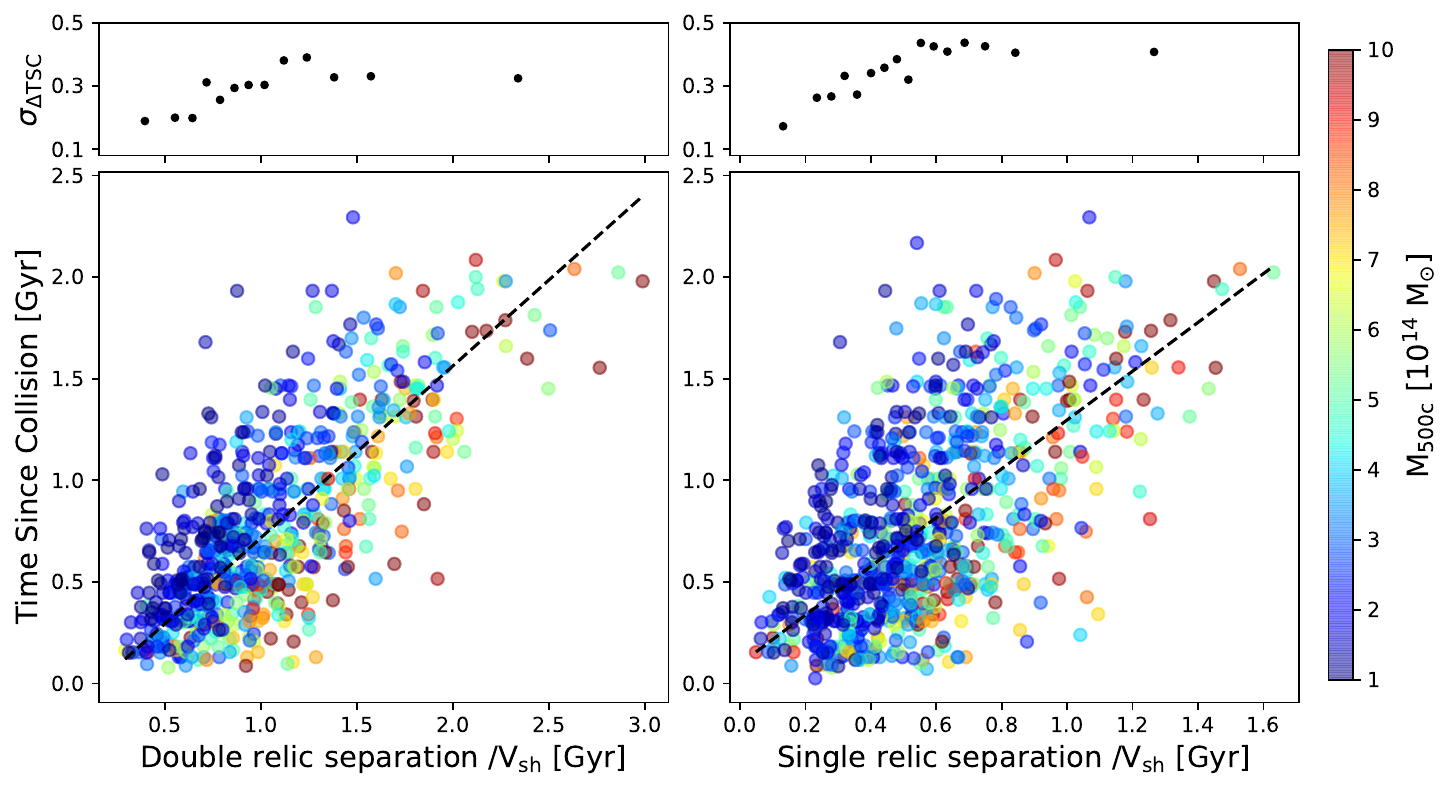}
 \caption{Same as Figure~\ref{fig:TSC_relic_sep}, but for the double (left) and single relic separations divided by the shock velocity $V_{\rm sh}$ (right). The shock velocity is estimated from the sound speed, derived using the mass–temperature relation, and the median Mach number of the relic cells. Colors indicate $M_{\rm 500c}$ of the clusters. The dashed lines show the best-fit linear relations over the ranges $[0.8, 2.0]$ and $[0.4, 1.0]$ for double- and single-relic systems, respectively. The time since collision correlates with the shock-inferred time, though the scatter is larger than in the $R_{\rm 500c}$-normalized correlations shown in Figure~\ref{fig:TSC_relic_sep}.
 }
 \label{fig: app_TSC_relic_sep}
\end{figure}

Figure~\ref{fig: app_TSC_relic_sep} compares the TSC and the shock-inferred time for double and single radio relic systems. Similar to Figure~\ref{fig:TSC_relic_sep}, the TSC correlates with the shock-inferred time. Fitting a linear relation over $d_{\rm drr}/V_{\rm sh} = [0.8,2.0]\rm~Gyr$ and $d_{\rm srr}/V_{\rm sh}=[0.4,1.0]\rm~Gyr$, yields $\rm{TSC} = 0.8 d_{\rm drr}/V_{\rm sh} - 0.1 \rm~Gyr$ and $\rm{TSC} = 1.2 d_{\rm srr}/V_{\rm sh}+0.1 \rm~Gyr$ for double and single relics, respectively.
The uncertainties of the TSC estimates range from 0.2 to 0.4 Gyr, which is comparable to or larger than those from the $R_{\rm 500c}$-normalized relations. The Pearson correlation coefficients are 0.62 and 0.35 for double- and single-relic systems, respectively, indicating a clear but weaker correlation than the relations discussed in \textsection\ref{sec: TSC}.

The larger scatter likely arises from the difference between the instantaneous shock strength and the cumulative evolution of the shock strength. In addition, normalization by $R_{\rm 500c}$ naturally incorporates the mass dependence encoded in the virial temperature.
Since merger shocks are comprised of shocks with a broad range of Mach numbers and as radio emission is more sensitive to the high Mach number regime (\textsection~\ref{sec: app_shock_property}), the shock-inferred time would be underestimated only using the radio observations. Thus, we conclude that $R_{\rm 500c}$ normalization of relic separation can achieve comparable or better accuracy than the shock-inferred time estimate, which further highlights the importance of accurate mass estimate.

\bibliography{reference}{}
\bibliographystyle{aasjournalv7}



\end{document}